\documentclass[10pt,english]{article}
\usepackage[letterpaper,bindingoffset=0.2in,%
            left=0.75in,right=0.75in,top=1in,bottom=1in,%
            footskip=.25in]{geometry}
\usepackage{cite}
\usepackage{notoccite}
\makeatletter
\renewcommand\@biblabel[1]{#1.}
\makeatother
\usepackage{amsthm}
\newlength{\bibitemsep}
\newlength{\bibparskip}
\let\oldthebibliography\thebibliography
\renewcommand\thebibliography[1]{%
  \oldthebibliography{#1}%
  \setlength{\parskip}{\bibitemsep}%
  \setlength{\itemsep}{\bibparskip}%
}

\usepackage{xcolor, setspace, graphicx, pict2e}
\usepackage{epstopdf}
\usepackage[normalem]{ulem}

\newsavebox{\ORCIDlogo}
\savebox{\ORCIDlogo}{%
\setlength{\unitlength}{\dimexpr 1em/256\relax}%
\begin{picture}(256,256)%
  \color[HTML]{A6CE39}\put(128,128){\circle*{256}}%
  \color{white}%
  \put(78.6,199.2){\circle*{20}}%
  \moveto(70.9,176,9)\lineto(86.3,176,9)\lineto(86.3,69.8)\lineto(70.9,69.8)%
  \closepath\fillpath%
  \moveto(108.9,176.9)\lineto(150.5,176.9)%
  \curveto(190.1,176.9)(207.5,148.6)(207.5 ,123.3)%
  \curveto(207.5,95,8)(186,69.7)(150.7,69.7)%
  \lineto(108.9,69.7)%
  \closepath\fillpath%
  \color[HTML]{A6CE39}%
  \moveto(124.3,83.6)\lineto(148.8,83.6)%
  \curveto(183.7,83.6)(191.7,110.1)(191.7,123.3)%
  \curveto(191.7,144.8)(178,163)(148,163)%
  \lineto(124.3,163)%
  \closepath\fillpath%
\end{picture}%
}

\newcommand\orcidicon[1]{\href{https://orcid.org/#1}{\usebox{\ORCIDlogo}}}

\usepackage[leftcaption]{sidecap}
\usepackage{lscape,amsmath,nccmath,nicefrac}
\newcommand\myeq{\mathrel{\overset{\makebox[0pt]{\mbox{\normalfont\tiny\sffamily def}}}{=}}}
\usepackage{relsize}
\usepackage{setspace}
 
\usepackage[T1]{fontenc}
\usepackage{footmisc}
\usepackage{hyperref} 
\hypersetup{colorlinks,breaklinks,
	unicode = true,
            urlcolor=[rgb]{0,0,.5},
            linkcolor=[rgb]{1,0,0},
            citecolor=[rgb]{0,0,1}}
\usepackage[all]{hypcap} 
\usepackage{caption}
\usepackage{makecell,multirow}
\usepackage{hhline}
\usepackage[normalem]{ulem} 
\usepackage{float}
\usepackage{centernot}
\usepackage{etoolbox}
\usepackage{amssymb}
\usepackage{tabularx}
\usepackage{wrapfig,booktabs}

\sidecaptionvpos{figure}{t}

\apptocmd{\lim}{\limits}{}{}
\newcolumntype{C}[1]{>{\centering\arraybackslash}p{#1}}

\newcommand{\cmmnt}[1]{\ignorespaces}
\newcolumntype{Y}{>{\centering\arraybackslash}X} 
\usepackage[euler]{textgreek}
\usepackage{cleveref}

\title{Reference standard analysis of multiple new and old plasma clearance models and renal clearance with special attention to measurement of reduced glomerular filtration rate \vspace{1.5cm}}

\author{{\bf Carl A. Wesolowski\,\orcidicon{0000-0003-0134-9346}$^{\dag\ddag\footnote{Corresponding Author: telephone :(306) 665 1515, e-mail: carl.wesolowski@gmail.com}}$} \vspace{0.5cm}\\
       $^{\dag}$  {\normalsize College of Pharmacy and Nutrition}\\
       {\normalsize University of Saskatchewan, 104 Clinic Place, Saskatoon, SK, S7N 2Z4, Canada}\\
       $^{\ddag}$ {\normalsize Department of Medical Imaging, Royal University Hospital, College of Medicine}\\
       {\normalsize University of Saskatchewan, 103 Hospital Drive, Saskatoon, SK, S7N 0W8, Canada}\vspace{11.5cm}}

\date{}
\begin{document}  


\begin{onehalfspacing}
\maketitle \vspace{0in} \noindent

\begin{abstract}
\noindent Nine models were evaluated as candidate glomerular filtration rate (GFR) reference standards in three datasets using  [$^{51}$Cr(EDTA)]$^-$ or [$^{169}$Yb(DTPA)]$^{2-}$ anions in 98 studies. Noncompartmental methods formed an upper limit for estimating mass excreted and voluntary urine collection formed a lower limit. For current models and methods, reduced GFR in adults resulted in inflated clearance estimates. Two different logarithmic models with exponential tails were created and may have underestimated reduced clearance. The logarithmic formulae can be used with only two plasma samples, and fit 13 studies totalling 162 plasma samples drawn from 5 min to 24 h with an 8\% standard deviation of residuals compared to 20\% error for monoexponentials. For shorter times (4 or 5 h) the fit errors decreased but the ratio of errors remained at circa 2.5 times lesser for the logarithmic versus monoexponential models. Adaptively regularised gamma variate, Tk-GV, models that are well documented, but not in common use, were largely contained within the reference extreme values, were unbiased for different levels of clearance and were the only models to be uncorrelated to volume of distribution from mean residence time divided by weight. Using Tk-GV as a candidate reference standard, potentially better methods for routine clinical usage were discussed. Prospective clinical testing, and metabolic scaling of decreased renal function is advised for potential changes to patient triage. 

\vspace{2em}

\noindent $\mathbf{Keywords}$: Glomerular Filtration Rate; Radiopharmaceuticals; Injections, Intravenous; Plasma; Reference Standards
\end{abstract}

\section*{Introduction}\label{introduction}
Glomerular filtration rate, GFR, can be  measured as the volume of arterial blood plasma per unit time totally cleared of nonindigenous, entirely-solvated, low-enough molecular-weight inert markers to be freely eliminated by renal filtration alone. GFR is widely considered to be the most useful measure of renal function \cite{Stevens2006}. This usefulness is likely due to a homeostatic balance between normal glomerular elimination of the products of metabolism and metabolic rate itself, such that reduced GFR signifies increased plasma concentration of a host of metabolites \cite{wesolowski2006improved}. This work presents and tests new and well known bolus intravenous GFR plasma models for use with venous sampling of radiochelates and other nonindigenous GFR markers for the purpose of stratifying models as to their relevance with respect to GFR reference standards. The bounds for reference standards used were noncompartmental plasma modelling and voluntary urinary drug mass collections. Moore \textit{et al}. found noncompartmental methods with an additional plasma volume concentration estimate at $t=0$ to overestimate renal clearance by circa 10\% \cite{moore2003conventional} at 4 h. Unfortunately, those authors did not test whether renal clearance should used as a reference standard. Most bolus intravenous injection pharmacokinetic models are venous plasma concentration sampling models of two principle types. The simplest and most commonly used type is the washout model; monotonically decreasing functions of time that have maximum concentration initially, at $t=0$. Models of the second type allow for the increasing concentration from an initial zero concentration in a peripheral sampling site, i.e., $C(0)=0$, and typically require more early data for fitting than  washout models. This work reports on several new washout models based on logarithmic functions having exponential tails, and a comparison of the results of multiple model types from three different series and two different radiopharmaceuticals.

\subsection*{The Schloerb challenge}\label{Schloerb}
 In 1960, Schloerb \cite{schloerb1960total} 
published the results of intravenous infusion of tritiated water, urea, and creatinine in nephrectomised dogs. Schloerb noted that plasma concentration of creatinine decreased with elapsing time and appeared to come to equilibrium after 4 hours, but then noted that this was only an apparent equilibrium as the expected complete equilibrium with total body water had not been achieved even at 24 h. He concluded that a near infinite number of compartments would need to be invoked to explain his results. That is, if we were to fit a monoexponential (E1) to Schloerb's disappearance curves, we would obtain a finite AUC, where AUC would have to be infinite to be consistent with the actual renal clearance of zero in a nephrectomised animal. Thus, monoexponentials and their sums fit to concentration curves from an infusion with data acquired for a short time exaggerate clearance. Moreover, most current models of plasma and renal clearance, be they from bolus intravenous injections, constant infusion, or subcutaneous injections do not reliably quantify renal insufficiency defined here as less than or equal to 25 ml/min for an adult. We refer to this problem as the Schloerb challenge, that is, to find a plasma disappearance curve model having a limiting infinite AUC with zero plasma clearance as renal clearance goes to zero.

Typical clinical measurements using monoexponential (E1) models collect two or more time-samples between 2 and 4 hours.  However, in severe renal insufficiency and/or fluid overload (ascites, tumour) the first time-sample should be collected at two or five h and the last at 24 h \cite{BroechnerMortensen1981,BroechnerMortensen1985,wickham2013development}, and even then the E1 results from 2 h to 24 h sample-times required correction for AUC underestimation \cite{wickham2013development}. One way to address the Schloerb challenge is to ignore plasma concentration models and instead measure GFR markers in urine. As Schloerb predicted, comparative measurements of E1 $\geq$ 2 h models of plasma clearance with renal (urine) clearance have shown that exponential plasma models predict substantial clearance values, when renal clearance was zero, i.e., causing an irreducible intercept error, e.g., 11.3 ml$\cdot$min$^{-1}$ \cite{LaFrance1988}. 

Current correction methods do not address the overestimation of zero renal clearance by plasma E1 models. For example, the Chantler-Barratt and Br{\o}chner-Mortensen, corrections of E1 clearance ($\text{CL}_{\text{E1}\geq2\,\text{h}}$) lack the appropriate nonlinearity at zero renal clearance to correct for a linear model's irreducible intercept, respectively, $\text{CL}\approx 0.87\, \text{CL}_{\text{E1}\geq2\,\text{h}}$ and $\text{CL}\approx 0.990778\, \text{CL}_{\text{E1}\geq2\,\text{h}}-0.001218\,\text{CL}_{\text{E1}\geq2\,\text{h}}^2$ \cite{Chantler1972,BroechnerMortensen1972,Murray2013}. Other formulas (Fleming, J{\o}dal, Ng \cite{Fleming2007,Joedal2009,Ng2011}) of the form  $\text{CL}\approx \text{CL}_{\text{E1}\geq2\,\text{h}}/(1-f\cdot\text{CL}_{\text{E1}\geq2\,\text{h}})$ are asymptotically $\text{CL}\simeq \text{CL}_{\text{E1}\geq2\,\text{h}}$ as clearance goes to zero, thus offer no correction for renal insufficiency. In specific, to reconcile a line equation negative intercept for using $\text{CL}_{\text{E1}\geq2\,\text{h}}$ plasma clearance to estimate renal clearance one requires a nonlinear equation with a slope at the origin that is asymptotically zero as in the contrary case, linear conversion risks returning negative numbers for low renal clearance values. Therefore, renal clearance is not being properly estimated, and it is clear that reference standards, including renal clearance, need to be investigated. 

A conversion of GFR to 1.73 m$^2$ divided by estimated body surface area (eBSA) is often performed. Although one can argue that creatinine plasma level scales approximately as BSA (circa weight to the 2/3 power), GFR certainly does not (circa weight to the 3/4 power)  \cite{adolph1949,wesolowski2006improved}. Another difficulty occurs in acute renal failure, which can be defined clinically by: creatinine levels (however, creatinine levels take days to build up); by loss of GFR, (presumably as GFR-indices from creatinine levels); or by 12 h of anuria or 24 h of severe oliguria of < 0.3 ml$\cdot$h$^{-1}$ per kg body weight \cite{Hilton2011}. In anuria, or severe oliguria, urine collection volumes are inadequate.  

This, and other factors, have led to a divergence between pharmacokinetics and nephrology with current nephrology guidelines suggesting multiple timed voluntary urine collections for a noisy \textit{underestimating} approximate body surface area normalised renal clearance reference standard from subcutaneous injections of ($^{125}$I)iothalamate, a marker with circa 18\% non-renal clearance \cite{Prueksaritanont1986}, see \nameref{Uprob} in the Methods section. That standard is currently recommended for calibrating a heuristic endogenous plasma creatinine GFR index \cite{delgado2022unifying}. Creatinine, in turn, is a mixed GFR and tubular extraction marker, and overestimates renal filtration in a variety of clinical conditions most notoriously in liver failure and renal insufficiency \cite{Wesolowski1992}. On the other hand, pharmacokinetics is concerned with drug effects most often correlated to venous plasma drug concentrations (GFR is arterial), utilise plasma (not renal) models that are tailored for route of administration, and might body scale per kilogram body mass for veterinary work, or occasionally BSA body scale for dose calculations, and would not likely claim that an 18\% non-renal cleared marker is a GFR marker. Thus, it is important to answer the Schloerb challenge as neither nephrologist nor pharmacokineticist has accurate methodology to offer the renal insufficient patient.

\subsection*{Answering the Schloerb challenge}\label{Sanswer}    Our first attempt to answer the Schloerb challenge produced the more accurate measurement of GFR obtained using the Tikhonov adaptively regularised gamma variate fitting (Tk-GV) method, which smooths the data to obtain that flattened curve that best reduces the relative error of propagation of the rate parameter of a gamma variate \cite{wesolowski2011validation,wickham2013development,wanasundara2016}. Because of this curve flattening, which becomes severe for renal failure, the Tk-GV algorithm is not a curve fit method in the ordinary sense. Compared to Tk-GV GFR-values, E1 and biexponential (E2) GFR values are larger, especially in severe renal insufficiency, because exponential methods overall underestimate both early and late concentrations \cite{wesolowski2010tikhonov,wesolowski2011validation,wickham2013development}. The use of the \text{Tk-GV} algorithm for measuring GFR was unique enough that patents were granted in the USA and multiple other jurisdictions \cite{wesolowski2014method}. 

For bolus intravenous injections, mixing takes a long time, thus concentration does not decrease in proportion to the logarithm of concentration. Indeed, in a prior publication, concentration before 2 to 4 h following a bolus injection of a GFR marker more accurately back-extrapolated as the logarithm of time, than as an area underestimating exponential, or an area overestimating power function \cite{wanasundara2015early}. The intent here was to characterise and test multiple models, and develop  bounds for GFR reference standards especially for reduced renal function.

\section*{Theory: The linear-logarithm hypothesis}

For a very long time it has been supposed that as a first approximation, the concentration of an intravenously injected GFR marker is proportional to the logarithm of concentration. That supposition implies an instantly achieved static volume of distribution with drug concentration that is changing in time. An additional requirement is sometime referred to as instant mixing, but strictly speaking the requirement is that the mean concentration within that volume is what is eliminated. In 2015, it was noted that during the first few hours following intravenous injections of a GFR marker, concentration decreased \textit{less} exponentially, i.e., \textit{less} linearly with the logarithm of \textit{concentration}, and decreased \textit{more} linearly with the logarithm of \textit{time} \cite{wanasundara2015early}. 

It would be better physiology to assume that early concentration is logarithmic as this assigns a starting volume of zero, but then modify the logarithm to later become exponential to allow for a terminal volume of drug distribution. In general, the family of functions having $t=0$ asymptotes that are logarithms and are asymptotic to zero concentration in the tail is the negative logarithm of sigmoid function family. Standard sigmoid functions have a slope of 1 at the origin and approach 1 from below in the right tail. Not all sigmoid functions are standard; some have slopes not equal to 1 at the origin. We examined two negative logarithmic sigmoid functions with exponential tails.\footnote{The two new formulas, LCE and ln-coth, are from a more general model $C(t)=c\ln \big(\frac{\alpha}{e^{\beta \,t}-1}+1\big).$ Setting $\alpha=1$ yields $-c\ln\left(1-e^{-\beta\, t}\right)$, which is the LCE function, and for 
$\alpha=2$, the general model reduces to $c\ln \big[\coth \big(\frac{\beta \, t}{2}\big)\big]$, the ln-coth model.} Of the many such formulas, one of them assigns concentration as proportional to $-\ln(1-e^{-\beta\,t})$, called the logarithm of cumulative exponential function (LCE), and another is $\ln\big[\coth(\frac{\beta\,t}{2})\big]$ called the ln-coth function. These functions correspond to model formulas that are presented in Table \ref{dists}, and whose derivations appear in the \nameref{sec:appendix} section. The LCE model is potentially the more useful one, such that more information is presented for it than for ln-coth. One can write LCE model in pharmacokinetic form using a constant of proportionality, $c=\text{AUC}\frac{6\, \beta }{\pi ^2}$,

\begin{equation}\label{eq4}
C(t)=-c\ln \left(1-e^{-\beta\, t}\right);\;\;\;
\text{AUC}=c\frac{\pi ^2}{6\, \beta }\;\;\;,
\end{equation}

\noindent called the LCE model as $1-e^{-\beta\, t}$ is the Cumulative Exponential distribution. Similarly, one can write the ln-coth pharmacokinetic model as,

\begin{equation}\label{coth}
C(t)=c\ln \bigg[\coth \bigg(\frac{\beta\, t}{2}\bigg)\bigg];\;\;\;
\text{AUC}=c\frac{\pi ^2}{4\, \beta }\;\;\;.
\end{equation}

\begin{table}[H]
\centering
\captionsetup{singlelinecheck=false,justification=raggedright,margin=0cm}
 \caption {Comparison of the ln-coth  and Logarithm of Cumulative Exponential (LCE) distributions. $^a$}
 \vspace*{-.5em}
\label{dists} 
\begin{tabularx}{\textwidth}{@{\hspace{0cm}}l@{\hspace{.4cm}}l@{\hspace{.4cm}}ll}
 \Xhline{2\arrayrulewidth} 

\vspace{0em}

Distribution&ln-coth&LCE&Notes\\\midrule
 \vspace{0em}
 
Type&Washout&Washout&Monotonic decreasing\\
\vspace{0em}

Parameters&$ \beta>0$, rate &$\beta>0$, rate & Rate is 1/scale\\
 \vspace{.4em}

Support&$t\in[0,\infty)$&$t\in[0,\infty)$&Semi-infinite support\\
 \vspace{0em}
 
Density function, $f(t)$&$\frac{4 \,\beta }{\pi ^2}\ln \left[\coth (\frac{\beta\, t}{2})\right]$&$-\frac{6\, \beta }{\pi ^2}\ln \left(1-e^{-\beta\, t}\right)$&Probability $f(t)$ only: PDF\\
 \vspace{0em}

CDF, $F(t)\;^\text{ b}$&$\frac{4 }{\pi ^2}\Big[\ln (y)\ln (y+1)$&$1-\frac{6 }{\pi ^2}\text{Li}_2\left(e^{-\beta\, t}\right)$&Li$_n(z)$ is the polylogarithm \\
&\hspace{0.8em}$+\text{Li}_2(1-y)$&&Li$_2(z)$ is a dilogarithm\\
&$\hspace{1.em}+\text{Li}_2(-y)\Big]+\frac{4}{3},$&&\\
& \& $y=\coth (\frac{ \beta \,t}{2})$&&\\
 \vspace{0em}

$t_{m}:F(t_m)=\frac{1}{2}$&$\approx \frac{0.526862}{\beta}$&$\approx \frac{0.415389}{\beta}$&Median residence time\\
 \vspace{.2em}
 
$\lim_{t\to0}f(t)$&$- \ln (\frac{\beta  \,t}{2})$&$-\ln (\beta \,t)$&Asymptotes logarithmic as $t \to 0$\\
  
$\lim_{t\to\infty}f(t)$&$2e^{-\beta  \,t}$&$e^{-\beta \,t}$&Asymptotes exponential at $t\to \infty$\\
\vspace{-.2em}
$t_{x}:$ limits $\equiv$ at &$\frac{W(2)}{\beta}\approx\frac{0.852606}{\beta}$&$\frac{W(1)}{\beta}=\frac{\Omega}{\beta}\approx\frac{0.567143}{\beta}$&Asymptotes intersect at $t_x$\\
\vspace{.2em}
&&&$\Omega$ is Lambert's $W(1)$\\
\vspace{.2em}

MRT $=\int_0^\infty t\,f(t)\,dt$&$\frac{7 \zeta (3)}{\pi ^2 \beta}\approx\frac{0.852557}{\beta}$&$\frac{6\, \zeta (3)}{\pi ^2\, \beta}\approx\frac{0.730763}{\beta }$&$\zeta (n)$ is the zeta function\\
 \vspace{.1em}

 V$_\text{MRT}=\text{CL MRT}$&$\frac{\text{CL}}{\beta}\frac{7 \zeta (3)}{\pi ^2 }$&$\frac{\text{CL}}{\beta}\frac{6\, \zeta (3)}{\pi ^2}$&\textit{Pharm.}: V$_{\text{SS}}$; Vol. steady state\\
 \vspace{.2em}
 
 $V_\text{d}(t)\;^\text{c}$&$0\leq\text{CL} \frac{1-F(t)}{f(t)}\leq\frac{\text{CL}}{\beta}$&$0\leq-\frac{\text{CL}}{\beta}\frac{\text{Li}_2\left(e^{-\beta \,t}\right)}{\ln \left(1-e^{-\beta \,t}\right)}\leq\frac{\text{CL}}{\beta}$&$V_\text{d}(0)\leq V_\text{d}(t)\leq V_\text{d}(\infty)$\\
  \vspace{.1em}

 $M_{\text{urine}}(t)$&$M_0 F(t)$&$M_0 F(t)$&Dose ($M_0$) in urine at time $t$\\
\hline
\end{tabularx}
\begin{tabularx}{1\textwidth}{X} 
 $^\text{a }$By definition a density function, $f(t)\myeq\frac{C(t)}{\text{AUC}}$, thus $C(t)=\text{AUC}\,f(t)$, also, see the \nameref{sec:appendix} section. \\
 $^\text{b }$CDF, the cumulative density function, is the integral of the density function, i.e., $F(t)=\int_0^t f(x)\,dx$.\\
 $^\text{c }V_d(t)$ for the ln-coth model is listed in unsubstituted (general) form  as its $F(t)$ is a long formula.
 \end{tabularx}
\end{table}

\begin{figure}[ht!]
\centering\includegraphics[scale=0.45]{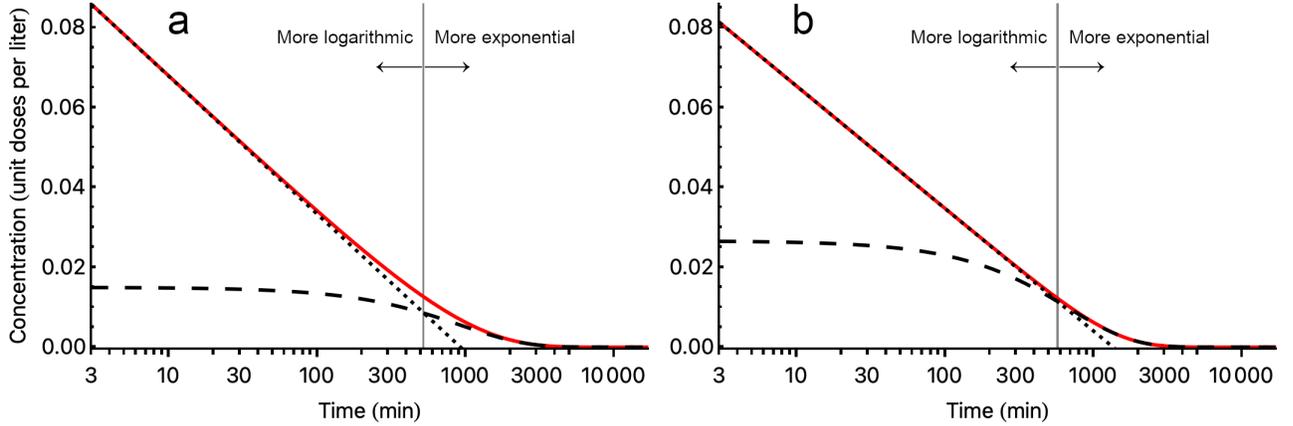}
\caption {Panel \textbf{a} shows an LCE model, $C(t)=-c\ln(1-e^{-\beta \,t})$, as a red coloured concentration versus time scaled logarithmically plot. Panel \textbf{b} shows an ln-coth model, $C(t)=c\ln \big[\coth \big(\frac{\beta\, t}{2}\big)\big]$, in red.  In both panels the logarithmic asymptotes are black and dotted, and the exponential asymptotes are black and dashed. For the ln-coth model, the intersection of its logarithmic and exponential functions are vertically closer to the model itself, i.e., the three curves shown overlap in panel \textbf{b} more closely than for the LCE model in panel \textbf{a}. From fits to the same time-samples, the intersection times, $t_x$, were similar but not identical, 523- and 576-min in panels \textbf{a} and \textbf{b}, respectively. }
\vspace{-1em}
\label{fig_1}
\end{figure}

As shown in Figure \ref{fig_1}, the LCE and ln-coth models, Eqs.~\eqref{eq4} and \eqref{coth}, each have two convergent asymptotes; the first a logarithm as $t\to0$ and the second an exponential as $t\to\infty$. There is a time when these asymptotes are equal, which for the LCE model is $\beta\, t$ such that,
$$-c\ln (\beta \,t)\equiv c\,e^{-\beta \,t}\;\;.$$  
Let $u=\beta\, t$, then as $c$ cancels, this equation becomes $-\ln (u)=e^{-u}$, whose solution is $u=\Omega$, where $\Omega$, is Lambert's Omega or $W(1)\approx0.567143$. Also called the product logarithm function, Lambert's $W(z)$, satisfies $w\, e^w=z$. In this case, $\Omega e^{\Omega}=1$, and we can write the intersection time for the asymptotes, $t_x$, as, 

$$t_{x}=\Omega\,\beta^{-1}\approx0.567143\,\beta^{-1}\;\;,$$

\noindent where $t_{x}$ is a time before which the LCE is predominantly a logarithmic function, and after which the LCE is relatively more exponential. From Table \ref{dists} and the \nameref{sec:appendix} section, the LCE model $t_m<t_{x}<\text{MRT}$. That is, the \textit{median} residence time ($t_{m}\approx 0.415389\,\beta^{-1}$) occurs when the LCE density is predominantly a logarithmic function of time, whereas its mean residence time (MRT $\approx 0.730763\,\beta^{-1}$), occurs when the LCE is more exponential. 

The intersection of the asymptotes of the ln-coth model occurs when $- \ln (\frac{\beta  \,t}{2})\equiv 2e^{-\beta  \,t}$, that is, at $W(2)/\beta$ (Table \ref{dists}). The ln-coth model has a more abrupt transition between its logarithmic and exponential asymptotes than the more gradually transitioning LCE model, see Figure \ref{fig_1}. The ln-coth model is a member of a larger family 

\noindent of functions; coth is hyperbolic cotangent, i.e., the reciprocal of hyperbolic tangent, and hyperbolic tangent is a standard sigmoid function; it goes through the origin with a slope of 1 and later approaches 1 from below. Any sigmoid function, $\textit{sf}\,(t)$, can be used to construct a terminal tail for a logarithm as $\lim_{t\to\infty}\ln[\frac{1}{\textit{sf}(t)}]\to0^+$, that is, as $\textit{sf}\,(t)$ approaches 1 from below $(1^-)$, its negative logarithm approaches zero from above ($0^+$), which causes concentration to be asymptotic to the late time axis. 

Other sigmoid functions, e.g., the error function, or the Gudermannian function could be used to make faster or slower decaying than exponential tails (\textit{stats}: lighter or heavier tails) in this same fashion. The LCE, ln-coth and Tk-GV models\footnote{where GV is a gamma variate; $C(t)=c\,t^{\alpha-1}e^{-\beta\,t}$, and the Tk-GV algorithm minimises the relative error of $\beta$.} (when $\alpha<1$) have zero initial volume of distribution, which requires an infinite concentration at $t=0$. For the Tk-GV model,  this is accomplished by adaptive fitting that yields $\alpha<1$. For all three models the infinity is \textit{integrable} and better mimics arterial concentration before the first sample times for small molecules like EDTA and DTPA chelates, and less so for inulin \cite{Cousins1997} and is our preferred method of adjusting venous sampling to arterial GFR conditions.

\begin{SCfigure}[][ht]\centering\includegraphics[scale=0.6]{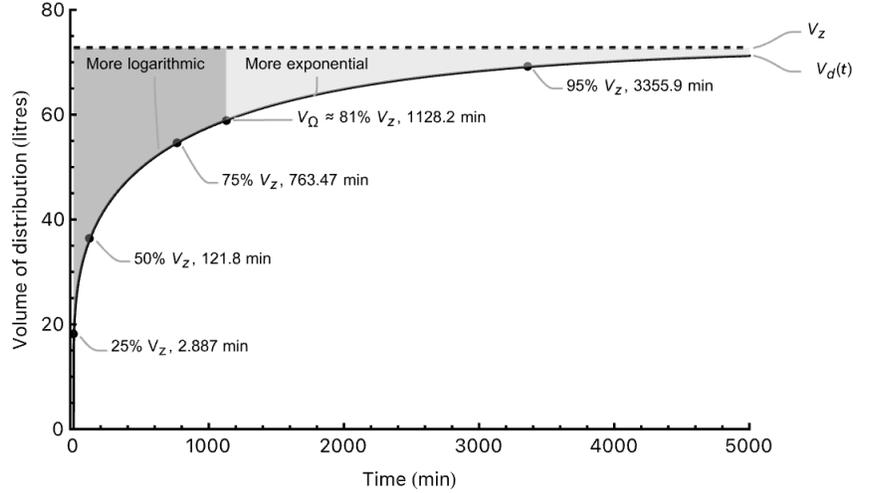}
\caption {Shown is a plot of LCE volume of distribution as a function of time, $V_d(t)$ (Table \ref{dists}), with reuse of the same parameters used to create Figure \ref{fig_1}. 
   Note that $V_\Omega \approx 0.81V_z $, where $V_\Omega $ occurs at $t_{x}=\Omega\,\beta^{-1}$.}
   \vspace*{-1em}
   \label{fig_2}
\end{SCfigure}

For $-$log-sigmoid models and sums of exponential term (SET) models the constants of proportionality are equal to the models' concentrations at different times. For SETs the total concentration $C(0)=c_1+c_2+c_3+\cdots+c_n$ at $t=0$. For the LCE  model the time when its  concentration equals $c$ occurs at $t:\ln(e-1)\,\beta^{-1}\approx 0.541325\,\beta^{-1}$. As per Table \ref{dists} and Figure \ref{fig_2}, the LCE and ln-coth models have a zero initial volume of distribution; $V_d(0)=0$, which is unlike the SET value, $V_c>0$, that is, the central (i.e., initial, non-zero) volume of distribution. For the LCE model, the  volume of drug distribution at which concentration curve shape becomes more exponential is 81\% $V_z$ occurring at time $t_x = \Omega\,\beta^{-1}$ and is a substantial portion of $V_z$, the terminal volume. This is from the LCE volume equation, $V_d(t)$, as follows,

\begin{equation}
V_\Omega=-\mfrac{\text{Li}_2\left(e^{-\Omega}\right)}{\ln \left(1-e^{-\Omega}\right)}\;V_z\approx 0.81000437\;V_z\;\;,
\end{equation}
where $V_\Omega$ is $V_d(t_x)$ and almost exactly 81\% of the LCE $V_z$.  For SETs, $V_c>0$, and $V_c$ is such that the mean concentration in that volume is assumed to be instantly presented for exchange between any compartments and sources of elimination. This unphysical assumption does not pertain to the Tk-GV, ln-coth and LCE models. The initial volumes of distribution are zero for the LCE, ln-coth and Tk-GV models, e.g., see  $V_d(t)$ in Table \ref{dists} and LCE in Figure \ref{fig_2}. 

\section*{Methods}\label{Methods}

\subsection*{Datasets 1-3}\label{S1-3}

\textbf{Dataset 1} was a group of 13 adult liver transplant candidates most having ascites who underwent bolus intravenous [$^{51}$Cr(EDTA)]$^-$ injections followed by plasma collection of a total of 162 time-samples drawn at 5 min to 24 h for routine assessment of renal function. Approval was obtained from the Royal Free Hospital Research Ethics Committee for the required extra blood sampling (REC reference number 07/H07211/70). The time-samples were obtained at circa 5, 10, 15, 20, 30, 40, 50, 60, 90, 120, 180, 240, 360, 480, 720, and 1440 min. The results of E1 and Tk-GV renal modelling appeared elsewhere \cite{wesolowski2011validation,wickham2013development}. 

\textbf{Dataset 2} was from  44 adults with cirrhosis and moderate to tense ascites from a project approved by the Ethics Committee for Medical Research in Copenhagen (J. nr. (KF) 11-110/02), i.e., group I of reference \cite{Henriksen2015}. These subjects underwent bolus [$^{51}$Cr(EDTA)]$^-$ intravenous injection followed by plasma collection of a total of 555 time-samples drawn at 5 min to 5 h, as well as circa 5 h of voluntary urine collection with assay of accumulated urinary drug activity. Time-samples were acquired at 0, 5, 10, 15, 30, 60, 90, 120, 150, 180, 240, and 300 min.

\textbf{Dataset 3} contains 328 plasma samples of [$^{169}$Yb(DTPA)]$^{2-}$ anion from 41 adult studies in whom time-samples were drawn at 10 min to 4 h following bolus intravenous injection. The eight time-samples in each study were collected at circa 10, 20, 30, 45, 60, 120, 180, and 240 min. These data are from an older study prior to routine publication of ethics committee identification numbers, but were nevertheless ethically obtained \cite{Russell1985}. At that time, there were problems with DTPA-chelate plasma binding \cite{Carlsen1980}, likely due to improper pH buffering in certain commercial DTPA chelation kits, and the [$^{169}$Yb(DTPA)]$^{2-}$ anion, gamma count time-samples were plasma protein binding corrected using ultrafiltration. This group had subjects whose renal function varied from renal failure to normal renal function \textit{without} evidence of fluid disturbance.

\subsection*{Urinary reference standards}\label{Uprob} 

Current nephrology guidelines recommend using a variation of voluntary urine collection data as a reference standard for calibration of GFR \cite{Meeusen2022}. Fortunately, the data here uses a better marker, [$^{51}$Cr(EDTA)]$^-$, and a better route of injection (intravenous) than the iothalamate and subcutaneous route\footnote{The subcutaneous route may have been chosen in an attempt to mimic constant infusion.} used for creatinine formula calibration. The classical renal clearance formula, used when constant infusion of a marker has reached a steady state plasma concentration, is CL is equal to $\frac{\text{U}\,\text{V}}{\text{P}}$, where U is Urinary concentration of an exogenous plasma marker during a short time interval, e.g., 20 min, some hours after infusion has begun, V is Volume of urine collected during that brief test time interval and P is the constant plasma concentration during that short collection time. Note that the product U$\times$V is marker mass accumulated during the urine collection. In their classical work, Walser and Bodenlos, using bolus intravenous E1 models, noted an unexpected 30 to 90 min delay between disappearance of radiolabeled urea from plasma and its appearance in urine \cite{walser1959urea}. This should serve as a reminder that $\text{CL}=\frac{\text{U V}}{\text{P}}$ is only defined for P (plasma concentration) under steady-state conditions. Dataset 2 lists total urinary drug mass (in our case radioactivity) collected during the entire circa 300 min following injection. This has the advantage of being more accurate in the sense of having a lot of data and not being a short collection time. However, the disadvantage of this is that the bolus intravenous plasma concentration curve changes in time, and is not any particular constant value, which prevents us from calculating a clearance without also knowing what the exact plasma concentration curve shape is. To be clear, each plasma concentration cumulative curve appropriate for use for a bolus experiment $\frac{\text{U V}}{\text{P}}$ calculation would be different  for each different curve model. It is possible to back calculate the renal CL-values for each plasma model, but that would not tell us which renal CL-value is correct. Accordingly, a different calculation was used for reference value testing. The objective of testing different plasma concentration curve models was accomplished by comparing the urinary drug mass collected (U V) with the mass predicted to be excreted from each plasma concentration bolus model ($M_0 F(t)$, Table \ref{dists}). Even then, there were further considerations.

The plasma concentration sampling time correction to account for the delay between zero time and marker first appearance in urine during a bolus experiment has been estimated as circa four min average, where literature estimates of average times were 2.5-8 min \cite{Ekins1966}. However, this time is longer in dilated urine collecting structures, e.g., renal pelvises and ureters, and for other reasons, e.g., renal insufficiency or intermittent obstructive disease. This time delay includes circulatory mixing time. That is, renal glomeruli filter arterial, not venous, blood. All of the plasma samples in this report are venous. Cousins \textit{et al.} showed negative arteriovenous differences for individual inulin and [$^{99m}$Tc(DTPA)$]^{2-}$ time-samples at 30 min and beyond\cite{Cousins1997}. Thus, the concentration appropriate as a divisor for the U V mass product, i.e., Urine drug concentration times Volume of urine, is a later, smaller, venous plasma concentration than the venous plasma concentration occurring at the time of urine collection with the effect that renal clearance will be otherwise underestimated. 

There are multiple other accuracy problems for voluntary urine collection: neglecting to save a voided volume \cite{John2016}; post void residual urine in the adult bladder \cite{Uensal2004}; worse and more variable residuals in the elderly from genitourinary pathology (including uterine prolapse and prostatic hypertrophy) \cite{Griffiths1996}; bladder resorption of x-ray contrast \cite{Currarino1977} and other drugs with resorption made worse with long elapsed time between voids \cite{Dalton1994,Wood1983}. Review of 24 h urine collections suggested that catheterisation avoids neglecting to save a voided volume and avoiding bladder drug resorption. Moreover, bladder catheterisation may correct some of the problems of residual urine in the bladder post void. However, even with catheterisation improper catheter placement itself led to residual bladder urine 26\% of the time \cite{Stoller1989}. Another problem is that there can be so little urine output in severe renal insufficiency that a small amount of bladder residual can render renal clearance based upon urine collection problematic.

In Dataset 2, case 6 of 44 had 6.5\% more urine mass collected (4.21$\times 10^{6}$ cpm) than administered (3.9533$\times 10^{6}$ cpm), which is unphysical. That case was excluded from mass balance comparisons. The other 43 cases were processed in two stages, initial screening, which showed an acceptable confidence interval agreement of mass balance between urine drug mass collected and the LCE and other methods of predicting urine drug mass. Subsequently, to test whether the agreement was only a statistical aberration, the LCE prediction was adjusted to occur four minutes earlier as per \cite{Ekins1966}, the voided volume was augmented by a positional average post void bladder residual of 13.014 ml as per \cite{Uensal2004}\footnote{13.014 ml is the straight average of five average residual bladder urine volumes from men and women after voiding in various positions.}  followed by discard of those voided volumes that were less than 70\% of predicted as recommended \cite{John2016}, wherein the frequency of incomplete urine collections was noted as 6\% to 47\%.  This procedure was repeated after dropping the initial time-samples to discover that LCE urine mass predictions from models whose first time-sample started at > 14 min agreed slightly better with the urinary mass calculations.

\subsection*{Noncompartmental reference standards}\label{NCprob} 

Noncompartmental exponential reference standards (NC) of clearance are often used by pharmacokineticists and were originally defined by Purves \cite{Purves1992}. This consisted of solving for the exponential functions that connected each adjacent plasma time-sample, then extrapolating using exponential fit functions to the last three or four samples, and when the concentration is increasing linear functions were recommended. For use here, the linear solutions and curve fitting were replaced with the solutions to the first or last sample and the weighted average of the next two or prior two samples. This provides two points, one natural and one averaged for an exact continuous solution that avoids having curve discontinuities at the extreme sample times. 

Consider, for example, that if at 300 min we had two different concentrations, one measured and one from a fit function, the urinary drug mass excreted at 300 min would be ambiguous. Solving for an extrapolating function that at 300 min has the same concentration as the time-sample itself obviates that problem, and works better. 

The formula for predicting drug mass (as cpm) excreted in urine at elapsed time, $t_U$, following bolus intravenous injection is approximately $M_U=\text{CL}\int_0^{t_U}C(t)\,dt$, where for noncompartmental (NC) methods, $C(t)$ is the piecewise defined concentration supported on $t=0$ to $\infty$ .

\subsection*{Summary of models used in this work}

Table \ref{mods} shows a summary of the models used in this work. Not all of the models were applied to all three datasets. In some cases, this is because they cannot be, for example the E1~$\geq 5$ h model proposed by Br{\o}chner-Mortensen and Freund \cite{BroechnerMortensen1981} can only be used for Dataset 1, which is the only one having enough temporal data for its application. Dataset 2 was particularly demanding as mass equivalent modelling was needed rather than renal clearance modelling. Renal clearance is best defined for steady state conditions following long term constant in-

\begin{table}[ht]
\centering
\captionsetup{singlelinecheck=false,justification=raggedright,margin=0cm}
 \caption {Summary of models used in this work}
 \vspace*{-.5em}
\label{mods} 
\begin{tabularx}{\textwidth}{@{\hspace{0cm}}l@{\hspace{.4cm}}c@{\hspace{.4cm}}ll}
 \Xhline{2\arrayrulewidth} 

\vspace{0em}

Model&$C(t)$&Description&Dataset $^a$\\\midrule
 \vspace{0em}
 
E1&$c\, e^{-\lambda\,t}$&Monoexponential&1, 2, 3\\
\vspace{0em}

E1 $\geq$ 2 h&"&E1 with time-samples $\geq$ 2 h& 1, 2, 3\\
 \vspace{.4em}

E1 $\geq$ 5 h&"&E1 with time-samples $\geq$ 5 h (24 h data  only)& 1\\
 \vspace{.4em}
 
E2&$c_1\,e^{-\lambda_1\,t}+c_2\,e^{-\lambda_2\,t}$&Biexponential& 1, 2, 3\\
 \vspace{.4em}

LCE&$-c\,\ln \left(1-e^{-\beta\, t}\right)$&Logarithm of cumulative exponential&1, 2, 3\\

LCE > 14 min&"&LCE with time-samples > 14 min&2\\

ln-coth&$c\,\ln \left[\coth (\frac{\beta\, t}{2})\right]$&Log hyperbolic cotangent&1, 2, 3\\

NC$^\text{ b}$&$-----$&Noncompartmental plasma model for excretion prediction&2\\

Tk-GV& $c\,t^{\alpha-1}e^{-\beta\,t}$&Tikhonov minimised relative error of $\beta$.&1, 2, 3\\

Urine&U$\cdot$V as (cpm/ml)$\cdot$(ml)&Drug mass (as cpm) in $\sim$300 min urine collection&2\\

\hline
\end{tabularx}
\begin{tabularx}{1\textwidth}{X} 
 $^\text{a }$For Dataset 2, the mass expected to be cleared is calculated at the end of the urine collection time with the exception of LCE > 14 min, which used a time 4 min earlier than that.\\
 $^\text{b }$See the \nameref{NCprob} section for the procedure.\\
 \end{tabularx}
\end{table}

\noindent fusion, not bolus intravenous conditions. The analysis for Dataset 2 includes three models not used elsewhere, (1) noncompartmental plasma model prediction of cumulative urinary drug mass (as radioactivity), (2) the adjusted LCE > 14 min excreted drug demonstration model, and (3) Urine, the total excreted drug mass calculation.

\subsection*{Statistical methods}

\subsubsection*{Regression analysis}\label{RA}

For each dataset several regression targets were tested for accuracy including: ordinary least squares (OLS), $\frac{1}{C_{obs}}$ weighted OLS, $\frac{1}{C_{obs}^2}$ weighted OLS, and OLS regression of log-log transformed $C_{obs}$ and sample times, where $C_{obs}$ are the observed concentrations. Of the regression targets tested, the $\frac{1}{C_{obs}^2}$ weighted OLS, also called proportional error modelling, proved the most accurate with the exception that log-log transformed regression is native to the Tk-GV clearance method, and not very different from proportional error modelling, see Eq.~(39) and surrounding text in reference \cite{wesolowski2020comparison}. For the Tk-GV method, the regression target is not curve fitting, but minimisation of the propagated proportional error of either clearance (CL) or of the exponential rate parameter ($\beta$) of a gamma distribution. Apart from the Tk-GV results, only the proportional minimum norm results are presented here. The regression method used for all targets was Nelder-Mead, which is more robust for absolute minimisation than gradient descent and most other methods, and is the most popular numericist's choice for regression analysis. Some pharmacokineticists prefer an adaptation of the maximum likelihood regression method from random variate minimisation, however, that was not tested here. The implementation was performed using the Mathematica 13.2.1.0 language on an Apple M1 iMac. All LCE model regressions converged rapidly, e.g., for Dataset 1 in 156.2 iterations at 52 milliseconds per case (mean values). For biexponentials, in one case of 57, the convergence was to a degenerate model, which 1.75\% failure rate is consistent with the circa 2\% failure rate reported elsewhere \cite{russell2002bayesian,wanasundara2016}. That model was $\lambda_2=\infty$ type; Dataset 2, case 19, 1470 iterations, 725 milliseconds, $C(t)= 0.100126 e^{-0.00755689 \,t}+0.0148127$, where $+0.0148127$ is a non-zero asymptotic value leading to CL$=0$. No other method yielded a zero CL for this case, the range being approximately 38.9 to 49.4 ml/min.

 Widely used for clinical laboratory assay calibration, Passing-Bablok type I linear regression was applied to the results including comparison of predicted and observed urine mass \cite{Bablok1983}. Passing-Bablok 
 type I regressions are used to evaluate replacement same-scale methods and are bivariate nonparametric regressions. In specific, these regressions find least squares in $x$ and $y$ where the regression target is replacement, that is, a best linear functional relationship, whereas ordinary (OLS) regression yields a minimum error line for predicting $y$-values. This is done to mitigate what for econometrics is called omitted variable bias for bivariate data, and for statistics is called regression dilution\cite{Clarke2005,Frost2000}. It corrects the flattening of slope (magnitude) that occurs when a least error predictor of $y$-alone, like ordinary least squares in $y$, is used to estimate a bivariate functional relationship, and is exaggerated for small magnitude correlations. Passing-Bablok regression works very accurately with good precision when comparing methods on the same scale, i.e., with slopes near 1, but it does so by discard of all possible two point negative slope combinations within the sample and then finding the median slope of the myriad combinations having positive slopes between any two points. Obviously, if the true slope were actually zero, Passing-Bablok would return a positive slope, so for slopes that are small in magnitude or negative the discards should not be performed. Passing-Bablok without negative slope discard is called Theil-Sen line regression and is both more robust to outliers and more accurate for bivariate problems than least squares in $y$, while not being completely unbiased for predicting bivariate linear relationships \cite{Wilcox1998}. Theil-Sen lines were used to examine how the differences between models behaved for various levels of renal function, for which the slopes can be zero or negative, i.e., Theil-Sen was used for those cases for which Passing-Bablok regression is not appropriate.
 
 \subsubsection*{Moving average and extrapolation testing}\label{average}
For residual analysis, i.e., of the difference between the concentrations of model values and time-samples, there is a need to examine how the models perform on average. As there are multiple plasma samples drawn at the same time following injection, one can take the number of earliest time-samples and average them to create a mean prediction for all the same model types. Next, one can drop an averaged time-sample from that group and bring in another averaged value from the next later group of time-samples, and assign that new group to have occurred at a new averaged time. This is performed until all the time samples have been average-averaged. This may seem contrived. However, if one were to drop and include unaveraged concentration values in each sample-time group, one would create a curve whose shape is dependant upon an arbitrary selection order of time-sample concentrations dropped or included. Finally, as each averaged, average-value is from the same number of averaged time-samples, it is equal-value weighted over the whole curve, and it is possible to do statistical analysis, such as finding a reliable standard deviation that shows how well model curve shapes match those of noise reduced data, and which procedure is asymptotically correct  as the number of samples increases. 

Extrapolation testing is done without withholding data by testing with Wilcoxon signed-rank sum one-sample differences from zero of the first and also the last groups of time-sample residuals from all of the curves in a dataset. Small probabilities indicate that it is \textit{unlikely} that the model extrapolates properly.   

\subsubsection*{Correlation of clearance to volume divided by weight}\label{V/W}
The reason for establishing that volume of distribution divided by weight is a relative constant irrespective of body habitus is because CL is spuriously correlated to volume of distribution via a third controlling variable: body mass. That is, mice with low body mass or smaller children have lesser clearance than elephants or larger children with larger body mass. It would appear that V/W is a normalisation that should be uncorrelated to clearance for a given population with certain exceptions. In ascites there is increased V/W, but within a given dataset of ascitic patients, there should still not be much if any covariation of V/W for CL. In renal failure it is possible to have increased sodium and body fluid for those patients who are not adequately controlled medically. This could lead to a negative correlation between CL and V/W. However, V/W > 1 as well as positive correlations between CL and V/W would not be so easily explained. 

As evidence that volume of distribution  of extracellular fluid\footnote{For plasma models, V is volume of drug distribution, not to be confused with the Volume of urine (also V) of a renal model.} (V) divided by weight (W) is a relative constant, we review a paper in which obese children were misleadingly claimed to have expanded extracellular fluid space compared to controls (Battistini 1995) \cite{Battistini1995}. This claim was made based on relatively reduced lean body mass for obese children, which as shown next is irrelevant. Those authors did not examine volume of distribution (V) by the bromine method divided by body mass (W) i.e., V/W. V/W in that paper was 12.3 litres for 56.8 kg obese children or  0.217 l/kg ($n=21$). For 18 controls, 8.9 litres corresponded to 41.0 kg body weight or also 0.217 l/kg. That is, there was no difference to three significant figures between values of V/W for obese versus control children.\footnote{Battistini \textit{et al.} used oral dosing of bromide, which is not as defensible as long term constant infusion, e.g., see Schwartz et al. \cite{Schwartz1949}, such that although their average of obese and normal V/W values are the same, both values may be underestimations.} As the density of human fat tissue is $0.9000 \pm 0.00068$ (mean $\pm$ standard deviation) \cite{Fidanza1953}, to make the same extracellular water content per kilogram as in denser tissues, there has to be less extracellular water per litre of fat, so there is in no sense expanded extracellular water content in fatty tissue. What there is, is relatively reduced intracellular water content in fat cells because water and fat are not very miscible. The authors neglected to appreciate that the ratio of V to ICW (intracellular water) increases not because V increases disproportionally (it does not, as above), but because ICW relatively decreases as relative fat content increases.

\section*{Results}%

\subsection*{Dataset 1 results}

\begin{figure}[H]
\centering
\includegraphics[scale=.44]{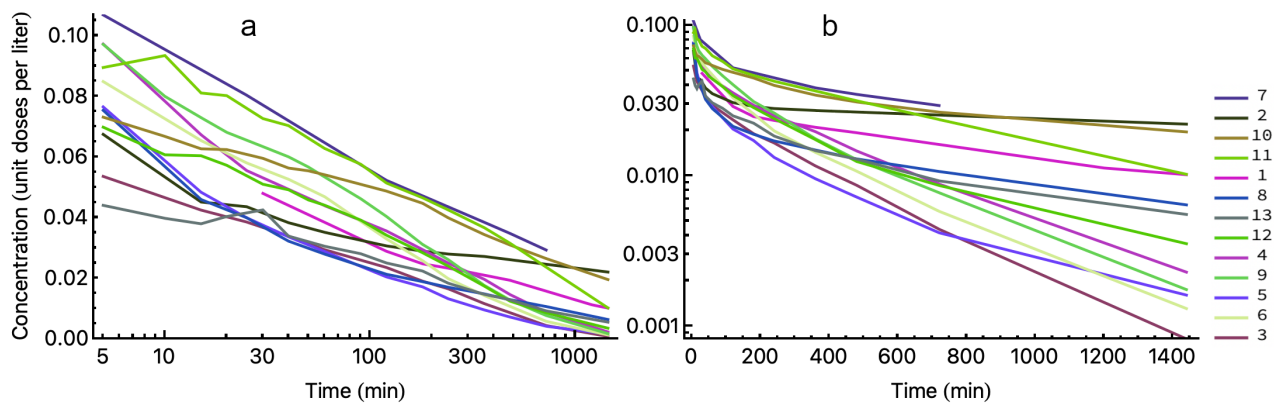}\captionsetup{belowskip=0pt}
\caption {Dataset 1 had 13 data series collected between 5 min and 24 h. These are shown as connected line segments, and plotted in two different ways. Panel \textbf{a} shows the cases plotted as linear concentration versus time on a logarithmic scale. Note the near linearity until late time of the line segments. Panel \textbf{b} shows semilog plots of the same data. Note the early time curvilinearity of the connected line segments.}
   \vspace{-1em}
   \label{fig_3}
\end{figure}

Figure \ref{fig_3} shows two competing plot types for viewing Dataset~1. The overall linear grouping of Figure \ref{fig_3}\textbf{a} can be interpreted as concentration propagating in time as a negative logarithm. However, negative logarithms would eventually yield negative concentrations. Thus, at some point in time, the logarithm should convert to an $x$-axis asymptote. Panel \textbf{b} shows relatively smooth but pronounced early-time log convexity,{\parfillskip=0pt\par} 

\begin{figure}[H]
\centering
\includegraphics[scale=.3105]{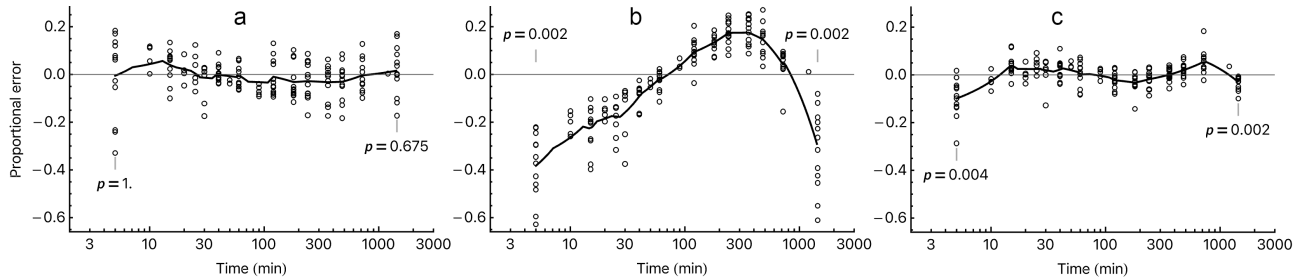}\captionsetup{belowskip=0pt}
\caption {Shown are Dataset 1 residuals for two parameter models in panel \textbf{a}; LCE models, and panel \textbf{b}; E1 models. Panel \textbf{c} shows four parameter biexponential model residuals. The circles are proportional modelling errors. The heavy black curves are 12 sample moving averages. The probabilities are the likelihood of the earliest and latest 12 samples having no fit error. The ln-coth fits are similar to panel \textbf{a}, see the text for the details.}
\label{fig_4}\vspace{-1em}
\end{figure}

\noindent   which are not linear and therefore not exponential for early-time on semi-log plotting. The curve fitting errors for those methods using proportional error modelling are displayed as residual plots in Figure \ref{fig_4}. Even though Dataset 1 has 13 cases, only 12 cases have 5 min time-samples and 12 have 24 h time-samples. A \textit{stationary} adaptation of a so-called \textit{moving} average of same sample-time averages was used as per the \nameref{average} Methods subsection. The standard deviation of those averages increased from a 1.83\% mean error of fitting of the ln-coth models, to a 2.38\% error for the LCE models, a 2.87\% error for the E2 models and a 14.17\% for the E1 models. For the LCE and ln-coth models, the 12 earliest and 12 latest time-sample errors were insignificantly different from zero, (respectively, $p=\{0.364,0.124\}$, and $p=\{1,\,0.675\}$) and very significantly different for the E1 and E2 models (respectively, $p=\{0.002,0.002\}$, and $p=\{0.004,0.002\}$).{\parfillskip=0pt\par}

\noindent  This suggests that on average for accuracy of curve fitting, ln-coth and LCE models  with only two parameters outperformed E1 and E2, despite the latter having an extra two fit parameters. The standard deviations of the residuals themselves worsen in a different order, 5.69\% for E2, 8.01\% for ln-coth, 8.42\% for LCE, and 20.07\% for E1.  Thus, the E2 fits, compared to the LCE and ln-coth fits are overfit, and overfitting can cause a spurious reduction of error under the curve, and does cause erroneous extrapolation \cite{Hawkins2004}, which given the significant earliest and latest time-sample underestimation causes underestimation of AUC and overestimation of CL. 

The results  in Table \ref{S1} shows the MRT values longer than the 24 h data (>1440 min) in bold font. The number of MRT-values longer than 24 h decreased in the following order: LCE, ln-coth, Tk-GV, E2, E1 having respectively 7, 4, 4, 2, 1 of 13 total.  The longer MRT-values led to larger AUC-values, and smaller clearances. The number of CL-values in the severe renal insufficiency range $(<20\text{ ml}\cdot\text{min}^{-1},$ bold type) decreased as LCE, ln-coth, Tk-GV, E2, E1 having respectively 5, 4, 3, 3, 1 of those CL-values. The smallest CL-value, LCE: 2.4 ml$\cdot$min$^{-1}$,{\parfillskip=0pt\par}

\begin{table}[H]
\centering
\captionsetup{singlelinecheck=false,justification=raggedright,margin=0cm}
 \caption {Dataset 1, some LCE, ln-coth, Tk-GV, biexponential (E2) and monoexponential (E1) model results.$^{\text{ a}}$}
 \vspace*{-.5em}
\label{S1} 
\small
\begin{tabularx}{\linewidth}{@{\hspace{.05cm}}l@{\hspace{.1cm}}c@{\hspace{.2cm}}c@{\hspace{.2cm}}c@{\hspace{.25cm}}c@{\hspace{.25cm}}cc@{\hspace{.2cm}}c@{\hspace{.1cm}}c@{\hspace{.1cm}}c@{\hspace{.3cm}}c@{\hspace{.3cm}}cc@{\hspace{.2cm}}c@{\hspace{.2cm}}c@{\hspace{.2cm}}c@{\hspace{.3cm}}c@{\hspace{.3cm}}c}
 \Xhline{2\arrayrulewidth} 
&\multicolumn{5}{c}{MRT (min)}& &\multicolumn{5}{c}{CL (ml$\cdot$min$^{-1}$)}& &\multicolumn{5}{c}{$V_\text{MRT}$ (L)}\\
&LCE&ln-coth&\scalebox{.8}[1.0]{Tk-GV}&\,E2&E1& &LCE&ln-coth&\scalebox{.8}[1.0]{Tk-GV}&\,E2&E1& &LCE&ln-coth&\scalebox{.8}[1.0]{Tk-GV}&\,E2&E1\\
\hhline{~-----~-----~-----}

Min&389&373&373&373&349& &\textbf{2.4}&\textbf{3.1}&\textbf{4.0}&\textbf{7.6}&\textbf{12.0}& &20.3&18.8&17.6&16.5&13.2\\
1st Quartile&451&431&453&437&365& &\textbf{14.4}&\textbf{18.7}&\textbf{18.7}&\textbf{18.8}&20.8& &24.8&21.8&20.7&20.3&17.1\\
Median&\textbf{1454}&1087&830&812&598& &36.6&40.5&37.5&40.2&45.9& &33.1&30.8&26.8&27.9&20.3\\
        3rd Quartile&\textbf{2873}&\textbf{1999}&\textbf{1995}&1189&863& &51.5&47.1&49.8&51.0&52.6& &51.6&42.2&39.9&35.3&30.5\\
    Max&\textbf{32735}&\textbf{20003}&\textbf{8395}&\textbf{4251}&\textbf{2285}& &84.4&81.6&79.7&80.3&86.6& &78.9&61.8&59.7&44.2&35.4\\
\hhline{~-----~-----~-----}
Mean&\textbf{4096}&\textbf{2662}&\textbf{1595}&1115&731& &37.3&38.3&37.6&39.7&42.7& &38.4&32.6&31.0&28.4&23.6\\
 \Xhline{2\arrayrulewidth} 
 \end{tabularx}
 \begin{tabularx}{1\textwidth}{X} 
 
 $^{\text{a }}$AUC is unit dose scaled. Results corresponding to MRT > 24 h and CL < 20 $\text{ml}\cdot\text{min}^{-1}$ are in \textbf{bold} font type.\\

\end{tabularx}
\vspace{-2em}
\end{table}

\noindent  had the longest MRT: 32735 min. The volumes of distribution (as $V_\text{MRT}$) decreased overall in the sequence LCE, ln-coth, Tk-GV, E2, E1. 

As mentioned in the Introduction, in severe renal insufficiency and/or fluid overload, there are two published suggestions for not using early time-samples to form better E1 model CL-prediction using 24 hours of data. The Wickham \textit{et al.} E1 $\geq$ 2 h method \cite{wickham2013development} would have us discard data before 2 h to improve CL-values overall, and the Br{\o}chner-Mortensen and Freund E1 $\geq$ 5 h method would have us discard data before 5 h to better predict severe renal insufficiency CL-values \cite{BroechnerMortensen1981}.  We compared proportional error regression for E1 models having time-samples $>0$, $\geq2$, or $\geq5$ h with the LCE and Tk-GV CL results. Table \ref{S1b} shows Passing-Bablok regression line prediction of the three CL$_{\text{E1}}$ models with the CL$_{\text{LCE}}$ and the CL$_{\text{Tk-GV}}$ values.  In that Table, as the earliest E1 data is increasingly ignored, the intercepts decrease in magnitude, but the slopes increase.{\parfillskip=0pt\par} 

\begin{wraptable}{r}{11cm}\small
\vspace{-0em}
\renewcommand{\arraystretch}{1} 
\linespread{1}\selectfont\centering
\caption{Dataset 1, Passing-Bablok regression line, $y=m\,x+b$, and confidence intervals (CI) of CL-values of LCE and Tk-GV ($x,\, \text{ml}\cdot\text{min}^{-1}$) versus E1 ($y$) models with various first time-samples, and correlations ($r$).}\label{S1b}
\vspace*{-2em}
\begin{tabular}{@{\hspace{0cm}}l@{\hspace{.2cm}}l@{\hspace{.2cm}}c@{\hspace{.1cm}}lc@{\hspace{.2cm}}lc}\\\toprule  
$x$,&$y$,&\hspace{0cm}$b$,&\hspace{.1em}   95\% CI $\left(\frac{\text{ml}}{\text{min}}\right)$&$m$,&\hspace{.1em}95\% CI&$r$\\\midrule
LCE&E1&\;\;\,11.63,&\;\;\;\,8.24 \;to 14.0&0.823,&0.709 to 0.992&0.97903\\ 
&E1 $\geq$ 2 h&\;\;\,6.886,&\;\;\;\,2.38 \;to 9.11&0.988,&0.903 to 1.103&0.99091\\
&E1 $\geq$ 5 h&\;\;\,4.362,&$-$0.327 to 6.65&1.144,&1.011 to 1.269&0.98635\\
Tk-GV&E1&\;\;\,7.386,& \;\;\;\;2.26\;\;to 9.75&0.899,&0.789 to 1.079& 0.97393\\
&E1 $\geq$ 2 h&\;\;\,1.813,&\;$-$2.34 \;to 4.86&1.113,&1.007 to 1.252&0.99049\\
&E1 $\geq$ 5 h&$-$2.410,&\;$-$7.07 \;to 2.18&1.303,&1.128 to 1.442&0.98723\\   \bottomrule
\vspace{-2em}
\end{tabular}
\end{wraptable}

\noindent   None of the E1 model types tested have both slopes of 1 and intercepts of 0 with confidence, which means that those E1 models are different from the LCE and Tk-GV models. Moreover, most of the intercepts are positive, which if true, means that to predict LCE or Tk-GV CL-values, negative intercept values would have to be subtracted from most of the E1 model types.\footnote{Note that the equations in Table \ref{S1b} can be solved for $x=m^*y+b^*$, where $m^*=1/m$ and $b^*=-b/m$, only because the regressions are Passing-Bablok type. In general, least squares in $y$ does not agree in that fashion with least squares in $x$.} Such intercepts are ill-conditioned as correction formulae because they may produce negative CL-values for reduced CL-values. To avoid negatives, E1 correction formulas should be non-linear, and go through the origin with slope zero at the origin when their Table 4 intercepts are positive.  

One quick way to check LCE and Tk-GV accuracy is to take their CL-values and divide that by the mean E1 CL, which yields a ratio of 0.873 for LCE and 0.879 for Tk-GV. Those ratios agree with the Chantler-Barratt \cite{Chantler1972} E1 correction factor of 0.87, so the LCE and Tk-GV mean CL-values, at least, are not implausible. However, as our objective was explore the entire range of CL-values with special attention to decreased renal function, it behoved us to do the same thing that Chantler and Barratt did, compare with urinary drug mass excreted. Thus, we next analysed Dataset 2, which has that information.

\subsection*{Dataset 2 results}

Table \;\ref{S2} \;shows \;Passing-Bablok

\begin{figure}[htb!]
\centering\includegraphics[scale=.575]{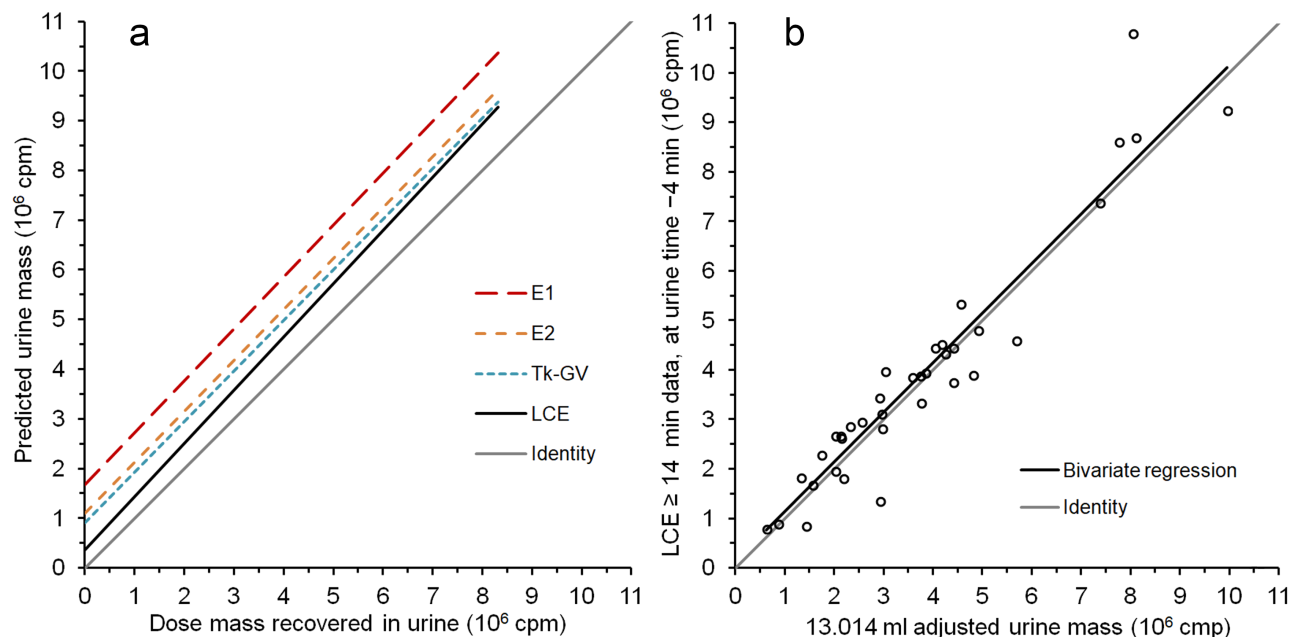}
\caption {As used for clinical laboratory assay calibration, Passing-Bablok type I regressions were used to evaluate equivalent or replacement same-scale methods for Dataset 2's voluntarily collected urine drug mass measured as 10$^6$ counts per min (cpm) of [$^{51}$Cr(EDTA)]$^-$ activity. Panel \textbf{a} shows mono- and bi-exponential (E1 \& E2), Tk-GV and LCE urinary mass predictions. Panel \textbf{b} shows bladder residual adjusted urine mass versus and LCE urinary mass predicted 4 min earlier from fits starting with > 14 min plasma data and then discard of 7 cases with less than 70\% of the LCE predicted urine drug mass. Only the LCE and LCE > 14 min models had no significant difference (95\% CI's) between slopes of 1 \textit{and} intercepts of 0 compared to urinary drug mass collected. The ln-coth model is not shown to avoid overlap, but appears in Table \ref{S2}.}
\label{fig_5}
\end{figure}

\begin{wraptable}{r}{12.0cm}
\small
\vspace{-3em}
\caption{Dataset 2, Passing-Bablok regression lines, $y=m\,x+b$, and confidence intervals (CI) for 8 models versus urine [$^{51}$Cr(EDTA)]$^-\,\times10^6$ cpm, number of cases (\textit{n}) and correlations ($r$).}
\label{S2}
\vspace*{-2em}
\linespread{1}\selectfont\centering
\begin{tabular}{@{\hspace{0cm}}l@{\hspace{.4cm}}c@{\hspace{.2cm}}lc@{\hspace{.2cm}}lcc}\\\toprule  
Urine $10^6\cdot$cpm,&$b$,&\;\;\;95\% CI&\hspace*{.3cm}$m$,

\;&95\% CI&\textit{n}&$r$\\\midrule
LCE > 14 min$^{\text{ a}}$&0.129,&$-$0.188 to 0.527&1.002,&0.893 to 1.119&36&0.95429\\
LCE&0.367,&$-$0.191 to 0.835&1.070,&0.903 to 1.232&43&0.90124\\
ln-coth&0.604,&\;\;\;0.147 to 1.258&1.107,&0.896 to 1.271&43&0.88820\\
NC&0.853,&\;\;\;0.487 to 1.548&1.036,&0.838 to 1.203&43&0.88431\\
Tk-GV&0.910,&\;\;\;0.402 to 1.337&1.018,&0.877 to 1.168&43&0.89411\\
E1 $\geq 2$ h&0.989&\;\;\;0.537 to 1.490&0.991&0.826 to 1.149&43&0.88570\\
E2&1.098,&\;\;\;0.624 to 1.649&1.027,&0.849 to 1.179&42&0.88507\\
E1&1.676,&\;\;\;0.820 to 2.065&1.046,&0.846 to 1.248&43&0.87096\\
 \bottomrule
 
\end{tabular}
\begin{tabularx}{1\textwidth}{X} 
 $^{\text{a }}$LCE  > 14 min was adjusted to 4 min earlier than the urine collection time. This\\
was the only model compared to $\sim$13 ml (residual) augmented urine volume (and\\ 
drug mass) with 7 cases discarded that had < 70\% predicted urinary drug mass.\\

\end{tabularx}
\end{wraptable}

\noindent  regression slopes and intercepts with 95\% confidence intervals for Dataset 2's 43 useful urinary masses at circa 300 min compared to the predicted amounts from 8 plasma models. Most of these regressed models appear in Figure \ref{fig_5}. Only the LCE and LCE > 14 min models had 95\% confidence intervals for intercepts that included zero, but all 8 plasma models had slopes that included one.

\begin{SCfigure}[][htb]
\centering\includegraphics[scale=0.7]{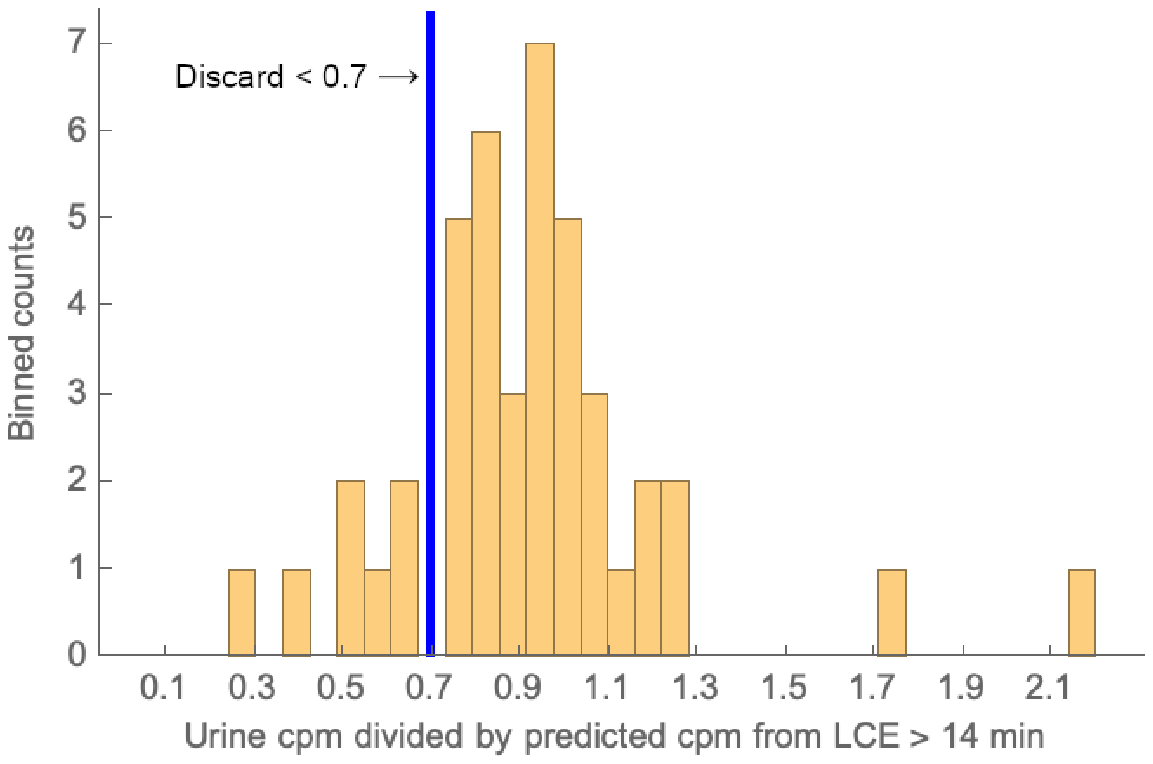}
\caption {Shown is a histogram of counts per minute (cpm), the radioactive equivalent of mass, in adjusted urine volume divided by LCE > 14 min models with predicted cpm at 4 min earlier than the end time of total urine collection. Note the blue line showing the discard upper limit of 0.7 used to create Figure 5b. Even though there are no values in the bin containing the 0.7 cut-point, there is no significant grouping into two populations: one with, and one without missed urine collection. The two apparent large ratio outliers may be due to underestimation of mass excreted by the LCE > 14 min models.}
\vspace{-2em}
\label{fig_6}
\end{SCfigure}

The LCE > 14 min model served to further demonstrate that the error between the LCE model and urine mass collected was negligible, with average error was reduced to 0.4\% by making multiple corrections. Those were by correction of urine count rate for 13.014 ml expected bladder residual, correction for a urine transit time of four min, by a slight improvement in the LCE fits by dropping early time-samples leaving for start time of > 14 min (LCE > 14 min), and finally by discard of the 7 recalculated urine samples with less than 70\% of the then adjusted LCE predicted activity to adjust for missing urine collections. This yielded tighter confidence intervals,
better correlation, and is illustrated in Figure \ref{fig_5}b. It is not known in absolute terms that the voluntary urine collections used here were incomplete \cite{Henriksen2015} and the literature is quite clear that a 70\% cutoff is heuristic \cite{John2016}. The histogram of ratios of corrected urine drug mass collected to corrected LCE > 14 min predicted mass of Figure~\ref{fig_6} has no strong evidence of two separate populations, but the sample size is small.{\parfillskip=0pt\par}


\noindent Nonetheless, one expects renal drug collection to be mass deficient at any particular elapsed time compared to pre-renal loss of drug mass at that same time due to numerous problems, as per the \nameref{Uprob} Methods subsection, including: possible missed collections of urine, urinary system transit delay time, possible bladder resorption of drug, possible increased urine dead spaces, and possible intermittent urinary obstruction. Thus, renal clearance appears to be a lower limit for reference clearance values. To complete the analysis, an upper limit for reference values was explored.

\begin{SCfigure}[][htb!]
\centering\includegraphics[scale=0.2]{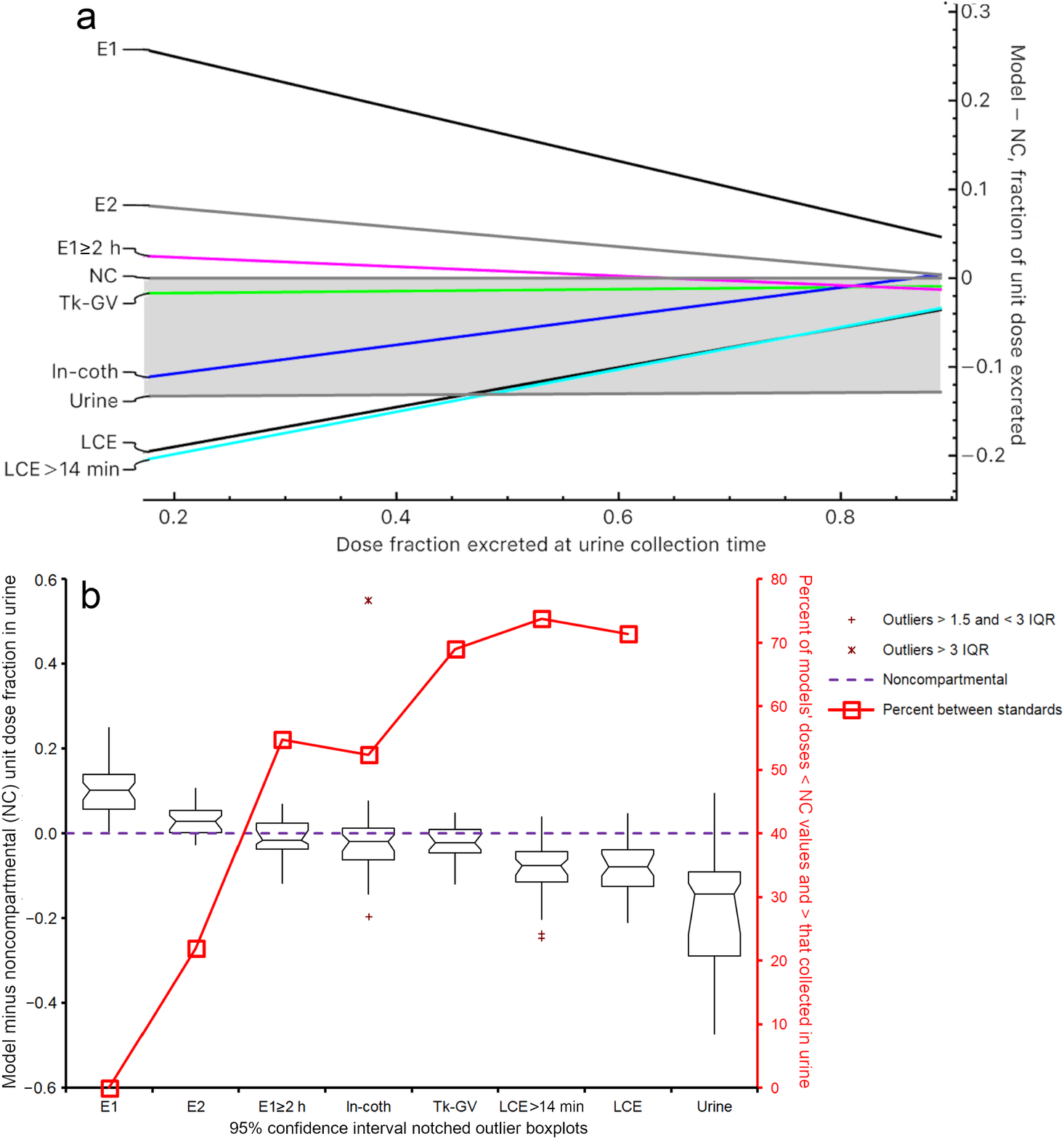}
\caption {Shown are various models' values minus each paired noncompartmental (NC) value. In panel \textbf{a}, Theil-Sen lines are shown from fits to those pair-wise differences. The grey area between NC estimated and (Urine) measured drug fraction illustrates the upper and lower bounds for a reference standard. Note that the curve fit methods are less accurate (with the exception of Tk-GV) when the mass excreted and clearance were reduced, with E1, E2 overestimating NC clearance. Panel \textbf{b} shows the non-parametric statistics as boxplots including median, quartiles, confidence intervals and outliers for each difference. The worst percentage of values within the bounds of the standard ranges were seen for E1 (0\%) and E2 (22\%), and was >50\% for all other fit functions. Note the increased variability of urinary drug mass collected minus NC drug mass excreted (Urine).}
\vspace{-1em}
\label{fig_7}
\end{SCfigure}

As we have seen for monoexponentials and biexponentials, the first and last time-samples are almost always underestimated concentration leading to overestimation of clearance. Consequently the  reference standard in common usage in pharmacokinetics, the plasma clearance noncompartmental method \cite{Purves1992}, that also uses exponential functions to extrapolate concentrations are, as per the methods used here, exact at the extreme sample times but still underestimate extrapolated and back extrapolated concentration as per \cite{moore2003conventional,wesolowski2010tikhonov,wanasundara2015early,Wesolowski2016PLoS}. Thus, the two standards in common usage; renal clearance and noncompartmental clearance, can be used to establish lower and upper respective bounds for reference standard values to then explore which if any of the other curve fitting methods examined produce results within those bounds. To explore this, the differences between NC and other model results were examined using Theil-Sen lines rather than Passing-Bablok regression, which latter is not useful for difference functions, see the \nameref{RA} Methods subsection. Figure \ref{fig_7}a shows Theil-Sen regression lines fit to the{\parfillskip=0pt\par}

\noindent models' predicted mass excreted with the noncompartmental paired values subtracted out. Using NC as a basis for this calculation rather than drug mass  in urine reduces noise, if for no other reason than plasma sample models are more alike to each other than any of them are to urinary drug mass measurements.  Interestingly, the Theil-Sen regression slope of the urine mass minus NC predicted mass excreted at that same time is minuscule (0.00653). Assuming that a proper reference method should be between the NC mass predicted and the measured drug mass at that time in urine, there are only two fit models' regression lines that fit that criterion, the Tk-GV and ln-coth models. Overall, the models performed worse for reduced renal function than for normal function as illustrated by their fan shaped divergent to the left of Figure \ref{fig_7}a. In the reduced function range the models ranked from overestimating to underestimating as  E1, E2, E1$\,\geq\,$2 h, NC, Tk-GV, ln-coth, urine mass, LCE and LCE$\,>\,$14 min. The E1 and E2 model lines did not cross into reference standard range at any level of function. Figure \ref{fig_7}b shows the sequentially decreasing median model minus NC pairs of the methods and the percent of values for each method included between the actual individual case values of the upper and lower reference standards. Of these, as likewise for Figure \ref{fig_7}a, the best fit function behaviour overall is from Tk-GV, having the least slope for a fit function (0.0109), the second least overall variability, best symmetry, and a good percentage of values within the  reference standard range (69\%). The LCE$\,>$14 min fit model had the largest percentage of values within the reference standard range at 74\%, followed by LCE fit to all time-samples at 71\%. 55\% of the  E1$\,\geq\,$2 h results were within the reference range.  

There were few results in the renal insufficiency range in Dataset 2, with the least plasma CL values for LCE, Tk-GV, E2 and E1 being 13.0, 24.3, 25.0 and 26.4 ml$\cdot$min$^{-1}$ respectively, with (uncorrected) LCE having three CL-values less than 20.0 ml$\cdot$min$^{-1}$. Tk-GV clearance was 3.33 ml$\cdot$min$^{-1}$ below the NC value (mean, \textit{p} = 0.0001, 2 sided \textit{t}-test) and its Passing-Bablok intercept was 5.47 ml$\cdot$min$^{-1}$ below the NC-value. The Chantler-Barratt style correction factor (OLS regression through origin for E1$\,\geq\,$2 h to predict LCE CL) for Dataset 2 was 0.781.

\subsection*{Dataset 3 results}\label{russel}

\begin{figure}[htbp!]
\centering
\includegraphics[scale=.46]{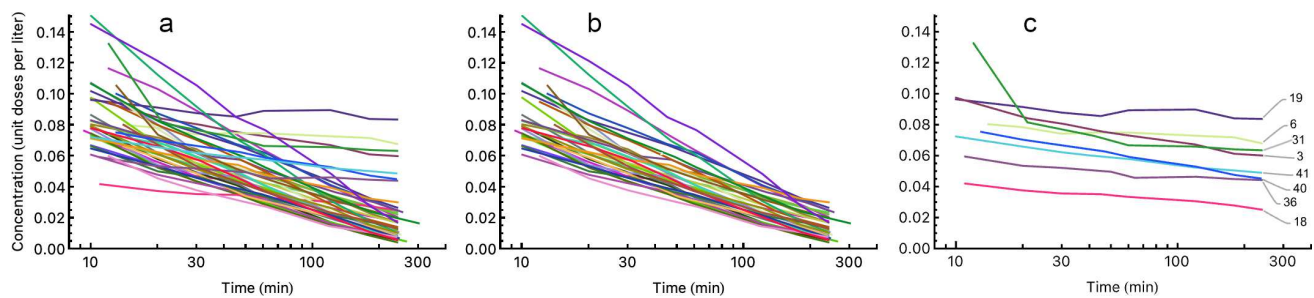}
\caption {Dataset 3 linear-log plots shown for all cases in panel \textbf{a}, without renal failure in panel \textbf{b} and with failure in panel \textbf{c}.}
\label{fig_8}
\vspace{-1em}
\end{figure}

Dataset 3 consists of 41 adult studies using [$^{169}$Yb(DTPA)]$^{2-}$ anion with eight time-samples collected from 10 min to 4 h.  Of interest for this dataset were how the formulae behaved 1) for a different GFR marker, 2) for subjects who did not have evidence of fluid disturbance and 3) for severe renal insufficiency. Upon LCE identification of nominal CL-values $<20\text{ ml}\cdot$min$^{-1}$, the dataset was sorted into cases with and without evidence of severe renal insufficiency. This is shown in Figure \ref{fig_8} as a clear difference between the behaviour of those two{\parfillskip=0pt\par}

\noindent  groups of studies.  That is, the suspected severely renal insufficient cases changed only slightly in concentration over 4 h, (Figure \ref{fig_8}c) as linearly decreased  concentration with elapsed time on a logarithmic scale. The more normal renal cases, Figure \ref{fig_8}b, approached the $t$-axis in late time as a  group, with sometimes slight asymptotic flattening in late time. The LCE model fit error for all 41 cases (Figure \ref{fig_8}a) was significantly greater than the fit error of the eight renal insufficient cases (4.88\%). The E1 models' error of fitting to these cases was 6.87\% and

\noindent significantly more variable than LCE fit error (Conover $p=0.003$). 

\begin{figure}[ht]
\vspace{.4em}
\centering
\includegraphics[scale=.55]{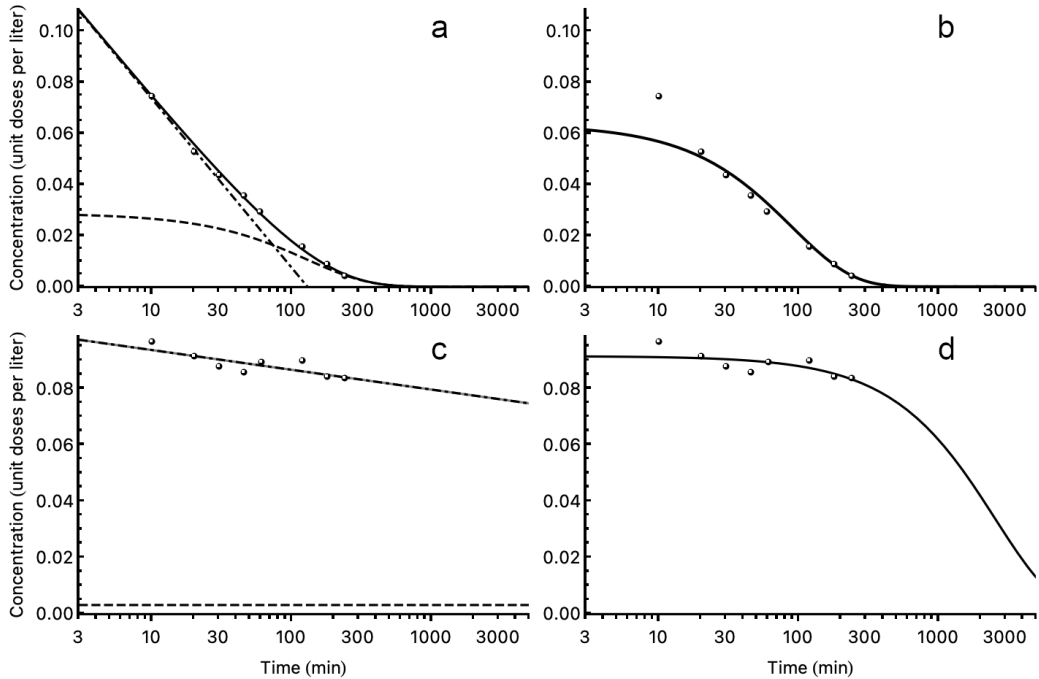} 
\vspace{-1em}
\caption {Dataset 3 linear-log plots of the greatest (case 15, panels \textbf{a} and \textbf{b}) and least (case 19, panels \textbf{c} and \textbf{d}) plasma CL-values from LCE and E1 models. Panel \textbf{a}, greatest CL LCE model. Panel \textbf{b}, greatest CL E1 model. Panel \textbf{c}, renal failure LCE model, and Panel \textbf{d}, renal failure E1 model. In panels \textbf{a} \& \textbf{c} the solid lines are the LCE models, the straight dot-dashed lines are the logarithmic early asymptotes and the dashed lines are the terminal exponentials. In Panel \textbf{c}, the early asymptote and model curve are superimposed.}
\label{fig_9}
\vspace{-1em}
\end{figure}

Figure \ref{fig_9} shows plots of the minimum and maximum plasma clearances cases for LCE and E1, where the LCE 

\noindent CL-values ranged from $9.27\times10^{-10}$ to 163.7 ml$\cdot$min$^{-1}$, and for E1 from 4.30 to 176.1 ml$\cdot$min$^{-1}$, respectively for cases 19 and 15. Overall, the fits for the LCE models have a 4.85\% standard deviation of proportional error, compared to 10.10\%  for E1. Note that these errors are approximately 1/2 of the values seen for Dataset 1, where Dataset 1 data was acquired for six times as long, i.e., 24 h versus 4 h. Figure \ref{fig_9}a shows an asymptotic approach to the time-axis after $t_x$, the intersection of the exponential curve and the early time asymptote; a straight line on linear-log plotting. However, in Figure \ref{fig_9}c, the LCE model and its logarithm are superimposed and the exponential (dashed) is flattened. In this worst case, the asymptotes intersected at a geologically long time; 4979 millennia. {\parfillskip=0pt\par}

\begin{wraptable}{r}{8.0cm}
\small
\vspace{-1em}
\caption{Dataset 3 renal failure candidates'  LCE, ln-coth, Tk-GV, E2 \& E1 model CLs (ml$\cdot\text{min}^{-1}$).}
\label{renfail}
\vspace*{-2em}
\linespread{1}\selectfont\centering
\begin{tabular}{@{\hspace{.1cm}}c@{\hspace{.1cm}}c@{\hspace{.1cm}}c@{\hspace{.2cm}}c@{\hspace{.3cm}}c@{\hspace{.5cm}}c}\\\toprule
Study N\textsuperscript{\underline{\scriptsize o}}&LCE&ln-coth&Tk-GV&E2&E1\\\midrule

19&9.27$\cdot10^{-10}$&1.24$\cdot10^{-9}$&1.24&2.60&4.30\\
6&1.19$\cdot10^{-6}$&1.58$\cdot10^{-6}$&2.85&5.56&7.05\\
36&0.0312&0.0416&6.29&5.63&18.2\\
41&0.406&0.504&10.0&11.7&22.3\\
3&1.06&1.41&9.49&13.9&20.3\\
31&1.13&1.48&5.72&8.30&17.3\\
18&2.89&3.83&27.2&43.5&48.7\\
40&3.17&4.18&17.0&20.7&30.0\\

\bottomrule
\vspace*{-2em}
\end{tabular}
\end{wraptable}

\noindent  In Table \ref{renfail} the largest LCE CL-value of 3.17 ml$\cdot\text{min}^{-1}$ for  these suspected renal failure cases had the shortest $t_x$ at 7.27 days; still largely beyond the capacity for validation for most experiments.  The E1 model only identified half of the eight severe renal insufficiency candidates of the LCE models. Proper identification of renal failure from E1 model usage is implausible as all 41 E1 models of Dataset 3 underestimated the concentrations of the first sample-times and 39 of 41 underestimated their last sample-time concentrations (Wilcoxon one-sample two-tailed $\textit{p}\ll0.0001$), and which correspond to systematic overestimation of CL, just as Schloerb observed. Similarly, the E2 first and last time-samples were significantly underestimating. The Tk-GV model identified seven of the eight cases having LCE CL $<20$, but at multiples of the LCE predicted plasma clearance values. The Chantler-Barratt style correction factor for Dataset 3 using LCE as the reference standard was 0.810, and 0.819 using Tk-GV.

\subsection*{Results, all datasets}
\begin{wrapfigure}{r}{0.5\textwidth}
 \vspace{-1em}
\begin{center}
 \includegraphics[scale=0.45]{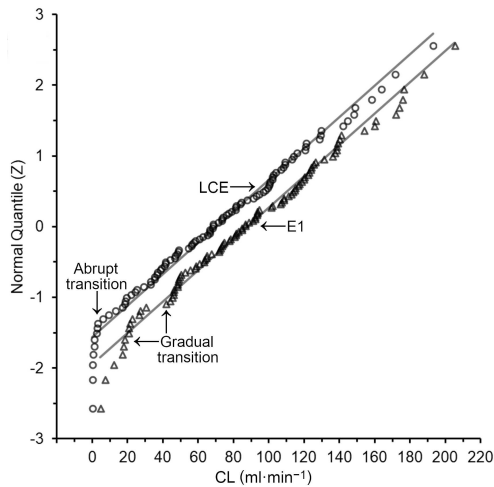}
 \vspace{-1.5em}
 \end{center}
\caption {Superimposed are the Q-Q plots for LEC (open circles) and E1 (open triangles) GFR-values. The solid grey lines give the locations of normally distributed values. The LCE values become abnormal very close to zero clearance. However, the E1 GFR-values transition beginning at approximately at 20 ml$\cdot$min$^{-1}$, which suggests why it is difficult to measure GFR < 20 ml$\cdot$min$^{-1}$ using current methods.}
\vspace{-1em}
\label{fig_10}
\end{wrapfigure}

For the total of 98 subjects analysed, there were 16, 13, 10, 9, and 6 having GFR-values < 20 ml$\cdot$min$^{-1}$ respectively for the LCE, ln-coth, Tk-GV, E2 and E1 models. The 95\% reference intervals for GFR were for: the LCE model from 0.015 to 167.9 ml$\cdot$min$^{-1}$; the ln-coth model from 0.020 to 172.7; the Tk-GV model 3.38 to 163.9; the E2 model 5.59 to 174.0, and for E1 9.40 to 182.2 ml$\cdot$min$^{-1}$, which explains the frequency of detection of the methods for GFR-values < 20 ml$\cdot$min$^{-1}$, e.g., E1 was unlikely to return a GFR value lower than 9.40 ml$\cdot$min$^{-1}$. Figure \ref{fig_10}{\parfillskip=0pt\par} 

\noindent shows how this occurred by quantile-quantile (Q-Q) plotting of all 98 GFR measurements for the LCE and the E1 models. This type of plot shows how measured values depart from the theoretical distribution used; in this case, the normal distribution. If one supposes that GFR-values are normally distributed a problem occurs because normal distributions extend from negative infinity to positive infinity, but GFR values cannot be less than zero. In practice that means that there should be a departure from normally distributed GFR-values in the region near zero GFR. Indeed, there is an abrupt departure from a normal distribution for the LCE model CL-values near zero, and a more gradual transition for the E1 CL-values. To investigate how abrupt this change should be the correlations between CL and fluid volume divided by weight, $\mfrac{V_\text{MRT}}{W}$, were examined, see Table \ref{correls}. Referring to that table, it is not obvious why the pattern of significance is different for dataset 2. The difference in pattern implies procedural or population differences between datasets such that all 98 studies were not correlation tested as a single group. Instead, a weighted average of correlations obtained in each dataset was used to rank correlations of each CL method with its V$_\text{MRT}/$W from greatest to least as E1, E2, Tk-GV, ln-coth, and LCE. {\parfillskip=0pt\par}

\begin{wraptable}{l}{6.8cm}
\small
\vspace{-1em}
\captionsetup{singlelinecheck=false,justification=raggedright}
\caption{Correlations of CL with V$_{\text{MRT}}$/W}
\label{correls}
\vspace*{-2em}
\linespread{1}\selectfont\centering
\begin{tabular}{@{\hspace{0cm}}l@{\hspace{.3cm}}c@{\hspace{.3cm}}c@{\hspace{.3cm}}c@{\hspace{.3cm}}c@{\hspace{0cm}}}\\\toprule
Dataset&1&2&3&all\\
$n$&13&44 (E2 43)&41&$n$-weighted\\
model&&&&mean\\
\midrule

E1&0.38&\textcolor{red}{\hspace{.7em}0.52} $^a$&0.27&0.40\\
E2&0.22&\textcolor{red}{0.33}&0.12&0.23\\
Tk-GV&-0.09&0.18&-0.01&0.06\\
ln-coth&\textcolor{red}{-0.57}&-0.01&\textcolor{red}{-0.53}&-0.30\\
LCE&\textcolor{red}{-0.67}&-0.15&\textcolor{red}{-0.54}&-0.38\\
  \hline
\multicolumn{5}{l}{$^\text{a }$ Significant results ($p<0.05$) in \textcolor{red}{red}.}
\vspace{-1em}
 \end{tabular}
\end{wraptable}

\noindent  For the three datasets, only Tk-GV had zero significant correlations. Taking at face value, it would seem that the Tk-GV models {\parfillskip=0pt\par}

\noindent yielded the more reliable volumes of drug distribution. As a further example, for Dataset 2 the noncompartmental reference standard CL-values were significantly correlated to its tediously calculated V$_\text{MRT}/$W, R = 0.36, with a 95\% confidence interval of 0.07 to 0.59, a significant result comparable to that of E2 models.

Both Kruskal-Wallis rank testing and 1-way ANOVA showed significantly different central measures of clearance between the hepatorenal compromised subjects in Dataset~1 (mean 37.3 ml$\cdot$min$^{-1}$) and the other datasets. However, there was no significant difference between Datasets 2 and 3 (means 73.6 and 77.0 ml$\cdot$min$^{-1}$, respectively), for LCE (or Tk-GV) CL-values from [$^{51}$Cr(EDTA)]$^-$ and [$^{169}$Yb(DTPA)]$^{2-}$ anions despite moderate to tense ascites in the former and the lack of fluid disturbance in the latter (Dataset 3): "Patients with edema ... were excluded from the study."\cite{Russell1985} Moreover, that clinical history can be examined retrospectively using the Tk-GV measures of V$_\text{MRT}/$W with 1-way ANOVA or the Kruskal-Wallis test, the results of which were in agreement. The ANOVA results are easier to follow. The mean Tk-GV V$_\text{MRT}/$W for Datasets 1, 2, and 3 were 0.386-, 0.293- and 0.248-l/kg, respectively. Normal extracellular fluid volume following 7.5 h (mean, $n=7$) constant infusion of thiocyanate anions was found to be 0.246 l/kg (mean) by Schwartz \textit{et al.}, Table I \cite{Schwartz1949}, such that the Dataset 3 Tk-GV almost identical mean of 0.248 l/kg seems normal range despite the methodological differences between studies.  However, by Dunnett contrasts, V$_\text{MRT}/$W was significantly increased in Datasets 1 and 2 compared to Dataset 3. In other words, there is no evidence of fluid disturbance in Dataset 3, whereas Datasets 1 and 2 have significant relative fluid disturbance. 

Seven of the ten Tk-GV CL-values less than 20 ml$\cdot$min$^{-1}$ were from Dataset 3, none were from Dataset 2 and three were from Dataset 1, such that if there were negative correlations between CL and V$_\text{MRT}/$W for Tk-GV values it would be seen in Datasets 1 and 3. There were small magnitude, insignificantly negative correlations from Datasets 1 and 3 between CL and V$_\text{MRT}/$W from Tk-GV processing, see Table \ref{correls}. Thus, one can say that the Tk-GV values for V$_\text{MRT}/$W are apparently consistent with the clinical history.  
On the other hand, LCE and ln-coth had significantly negative correlations between reduced CL and V$_\text{MRT}/$W for Datasets 1 and 3, with some reduced CL-values for V$_\text{MRT}/$W  that were > 1. That type of physiologic behaviour cannot be ruled out with certainty, but at face value seems less plausible than the results from Tk-GV.

Finally, the Chantler-Barratt style E1$\,\geq\,$2 h correction factor using the LCE model as the standard for all 98 cases was 0.800, and for TK-GV CL-values was 0.824.

\section*{Discussion}


\noindent The initial concentration in a peripheral venous sampling site is zero at the time of a bolus intravenous injection in a different vein.  To model the entire concentration curve including the zero initial concentration typically requires more early data, processing and theory than are typically used for routine drug assessment \cite{Wesolowski2016GDC,wesolowski2020comparison}. The alternative is to use incomplete models that do not model the very rapidly changing early vascular effects with the caveat that first time-sampling be drawn some minutes or hours following the time of peak venous concentration.  How many minutes or hours following injection one should wait to take samples depends on the model. For the Tk-GV model, 5 min is enough. For bolus injections of inulin and [$^{99m}$Tc(DTPA)]$^{2-}$ anions, 25 minutes in adult humans was the time at which arteriovenous differences of concentration concentrations equalised \cite{Cousins1997}. The LCE model produced possibly slightly better results with sampling times starting at 15 min rather than 5 min for Dataset 2, compared to start times beginning at 2 or 5 hours and ongoing for 24 h for E1 as suggested by Wickham \textit{et al.} \cite{wickham2013development} and Br{\o}chner-Mortensen \cite{BroechnerMortensen1981}, respectively. Table \ref{S1b} shows this effect for Dataset 1, the only dataset with 24 h data. Compared to the LCE and Tk-GV model CL-values, using an E1 model with 24 h data beginning at 2- or 5-h proved more accurate than fitting E1 to the complete data beginning at 5 min, but this comes at the cost of having to acquire 24 h data and still having to use correction formulas (Chantler-Barratt,  Br{\o}chner-Mortensen, and other corrections).   

Not unexpectedly,  the results showed that the attempts to fit E1 or E2 to time-limited data resulted in poor quality fits of the AUC underestimating type attributed to the curve shape of the data being more linear-logarithmic than exponential. This was the same problem for all three datasets, and is shown for Dataset 1 in Figure \ref{fig_4}. The change in concentration as apportioned in time logarithmically is not unknown. For instance, in Datasets 1 and 3 above, the time-samples were independently selected to be drawn at times that form a nearly geometric progression, where for example, a perfectly geometric progression would be a doubling time: 7.5, 15, 30, 60, 120, 240, 480,$\dots$ min. Such a scale is equidistant when its logarithm is taken, where the motive for doing so is to acquire data such that the change in concentration is more or less linear and detectable between time-samples. So clearly equal log-time, time-sample spacing is appreciated by some experimentalists. The search for incorporating that observation into a plausible model that forms a better basis for quantifying concentration curves than exponentials yielded several models, and potentially many others. For a more general model, $C(t)=c\ln \big(\frac{\alpha}{e^{\beta \,t}-1}+1\big)$, Lambert's $W$ solves $t_x=\frac{W(\alpha)}{\beta}$ as the time at which the asymptotes are equal. For example, the LCE model results from setting $\alpha=1$, and the ln-coth model results when $\alpha=2$. In even more general terms, the asymptotes of the negative logarithms of sigmoid functions may not intersect at all. For example, $-c\ln[\text{erf}(\beta\,t)]$, where erf is the error function, has a tail whose decay is so fast (\textit{stats}, light) that an intersection of its asymptotes\footnote{\hspace{.1em}The asymptotes are $c\ln \left(\frac{\sqrt{\pi }}{2 \beta  \,t}\right)$ and $-c\ln \left(1-\frac{e^{-(\beta \,t)^2}}{\sqrt{\pi } \beta \,t}\right)$} does not occur. However, even in that case, there is a local minimum concentration difference between those asymptotes that signals when the character of the curves changes from its logarithmic predominant shape to its tail shape.

The behaviour of negative log-sigmoid functions is every bit as complicated as that of biexponentials. For a biexponential, without loss of generality, one assumes $\lambda_1>\lambda_2$, and there are two compartments:  A central compartment with concentration $C_C(t)=c_1e^{-\lambda_1 t}+c_2e^{-\lambda_2 t}$, and a peripheral compartment with concentration $C_P(t)=\frac{c_1\, \lambda_2+c_2 \,\lambda_1}{\lambda_1-\lambda_2}(e^{-\lambda_2 t}-e^{-\lambda_1 t})$. The time at which those concentration curves are equal is $\frac{\ln \left(\lambda_1/\lambda_2 \right)}{\lambda_1 -\lambda_2 }$. Sometimes (if rarely) called the time of pseudoequilibrium, that is also the time at which $C_P$ concentration peaks before which is called the \textit{distribution phase}, and after which is called the \textit{elimination phase}. So, recapping, the $-$log-sigmoid functions have  distinct curve phases just like biexponentials do, but have a larger selection of tail behaviours, whereas biexponentials have twice the number of parameters for little gain in goodness of fit.

Biexponentials, and higher order mammalian models have compartments that imply a diffusion or osmotic barrier type of exchange predicated upon flow being proportional to a concentration difference across semipermeable membranes (forward osmosis), whereas many capillary membranes are washed, and undergo some combination of reverse osmotic and bulk pore flow with solutes being carried away on both membrane sides such that as a first approximation GFR markers are transported physically due to pressure differences, not concentration differences. In the kidney this is called ultrafiltration. To put it another way, when clearance is held constant, we typically assume that renal filtration of a solute is proportional to concentration of that solute, and as much of the small molecule transport to the interstitium occurs in those capillaries that have similar architectures, pressure differences and functionality \cite{Rostgaard1997}, there is no special call for diffusion barrier modelling, at least not for GFR markers. Schwartz \textit{et al.} (1949) \cite{Schwartz1949} thought that inulin (circa 3,500$-$5,500 Da) was a better extracellular fluid marker than thiocyanate or bromide based on the assumption that diffusion into intracellular water was occurring for the smaller molecules. The concept of molecular sieving through capillary pores dates from Pappenheimer \textit{et al.} (1951) \cite{Pappenheimer1951,Pappenheimer1953} and provides an alternative explanation. That concept implies that inulin's volume of distribution is smaller because the molecule is so large that its flow through capillary pores is partly impeded, which would not be the case smaller molecules like EDTA and DTPA chelates, thiocyanate and bromide. Electron microscopic examination of renal glomerular and other capillaries has demonstrated the existence of these high flow rate pores in some but not all tissues. \cite{Rostgaard1997} Finally, those tissues without high flow rate pores still have reverse osmosis as an anion transport mechanism.

An alternative explanation for biexponential behaviour with greater generality does not invoke compartments. That is, the variable volume model as first proposed by Niazi and later extended to all concentration scaled semi-infinite density functions \cite{niazi1976volume,Wesolowski2016PLoS}. Indeed, examination of how drug volume changes in time shows that biexponential and other summed exponentials all have a large initial volume of distribution which is unphysical, and is unlike Figure \ref{fig_2}, which is a variable volume of drug distribution plot that starts with zero initial drug volume by virtue of have an unbounded large initial concentration from an LCE model. It is possible to extend biexponentials and other washout models to have zero initial concentration, and zero initial drug volume by convolution modelling, but that is not a feature of simple washout models. George Box, a statistician, is well known for having written "Essentially, all models are wrong, but some are useful" \cite{box1987empirical}. Indeed, sums of exponential term models are used almost to the exclusion of everything else. However, E2 models are all too frequently unphysical, and was the only model that was not robust when applied to the data. For example, Dataset 2 subject 19 had zero E2 clearance, with other models having 38-49 ml/min. Finally E2 models were outperformed by noncompartmental methods.

The development of clinical reference standards is important \cite{Schimmel2015}. There is a need to refine reference standards for measured GFR. All too frequently, a GFR reference standard is assumed without preamble to be a true gold standard. For example, one of the E1$\,\geq\,$2~h clearances correction papers cited above used the word \textit{true} to describe an E2 model no less than two dozen times, the same model shown here to produce inflated CL-values compared to noncompartmental methods. In turn, noncompartmental methods yielded inflated CL-values compared to renal clearance, which latter some authors \textit{assume} to be true clearance \cite{moore2003conventional}.  Measurement standards evolve only through extensive testing. Gone, for example, is the circa 1901 platinum-iridium standard reference kilogram \cite{Richard2016}. Finally in 2019, following an effort lasting four decades, one kilogram was redefined as equal to the speed of light squared divided by the product of Planck's constant and the hyperfine transition frequency of $^{133}$Cs, and is precise and accurate to within several parts per billion. 
Compared to that, the median difference between the NC and urinary drug standard of 14 parts per hundred seems imprecise.

 Of the many tests performed in this paper several stand out as critical to our understanding of what a plausible reference standard for measured GFR should be. An important result was the test for correlation of CL with volume of distribution divided by weight, Table \ref{correls}. An unanticipated outcome was that the least correlated results were for the Tk-GV models, which were without significant correlation for all three datasets. This reflects the assumption that CL and the drug volume of distribution should be largely uncorrelated. In that same vein, Figure \ref{fig_7} is particularly revealing. In Figure \ref{fig_7}a, we noted that many of the eight models compared to noncompartmental models had more error measuring decreased cleared mass than the more normal range values did. This is an increased error of absolute drug mass cleared, and not just a relative or percentage value. That result implicates reduced clearance as meriting special attention for the evaluation of reference standards, and that clinicians should be aware that reduced GFR measurements obtained from most current methods tend to underestimate the severity of the reduction. Of the eight methods  compared to noncompartmental methods in Figure \ref{fig_7}, only two were not slope biased on the lower end of renal function. Those two were the Tk-GV plasma clearance method, and the renal clearance method. The most moderate of these methods, and the most plausible was the Tk-GV method, and that is unfortunate because it is not widely available, and those interested in using it should contact the author. Compared to the NC method the Tk-GV results were the second least variable (Figure \ref{fig_7}b). That is not too surprising because the least variable (by a hair) was the E2 model, which structurally is closely related the NC reference standard to which all others models were compared in that figure. However, the E2 models yielded so few results (22\%) between those of the NC method and actual urine mass of drug collected that they are not plausibly accurate measurements. The most frequently seen results within the reference range, 74\%, were from LCE fits starting at > 14 min.

\subsection*{Discussion, clinical relevance}

Using current methods, few measured plasma and renal GFR clinical studies are performed for patients having less than 20 ml$\cdot$min$^{-1}$,  e.g., there were none in Dataset 2. Renal clearance was well emulated by plasma CL$_{\text{LCE}\,>\,14\;\text{min}}$,  However, neither measure included reduced CL-values and a patient having 10 ml$\cdot$min$^{-1}$ Tk-GV clearance may merit different management than one with 0.406 to 22.3 ml$\cdot$min$^{-1}$, i.e., the range of CL-values in Table \ref{renfail} of study 41.  In lieu of a direct measurement of GFR, a current practice is, for example, to use the average of creatinine and urea renal CL-values or 24 h creatinine renal CL-values as well as urinary albumin levels as rough indicators of what the appropriate clinical management may be \cite{BroechnerMortensen1981}. Even using exogenous radiotracers, bolus injection urinary collection measurements are problematic, see the \nameref{Uprob} Methods subsection for details. For example, an oliguric patient may have undetectable renal clearance values. In prior work, Tk-GV clearances were more accurate and precise than E2 clearances \cite{wanasundara2016}. Most current plasma clearance methods fail the Schloerb challenge to quantify a lack of renal function but the TK-GV method apparently succeeded. However, only prospective studies can determine how a method agrees with other patient  management indicators, including selecting patients for dialysis, or reliably detecting even moderate loss of renal function from chemotherapy, radiation therapy, surgery, or disease.  

\begin{figure}[ht!]
\vspace{-1em}
\centering
\includegraphics[scale=.242]{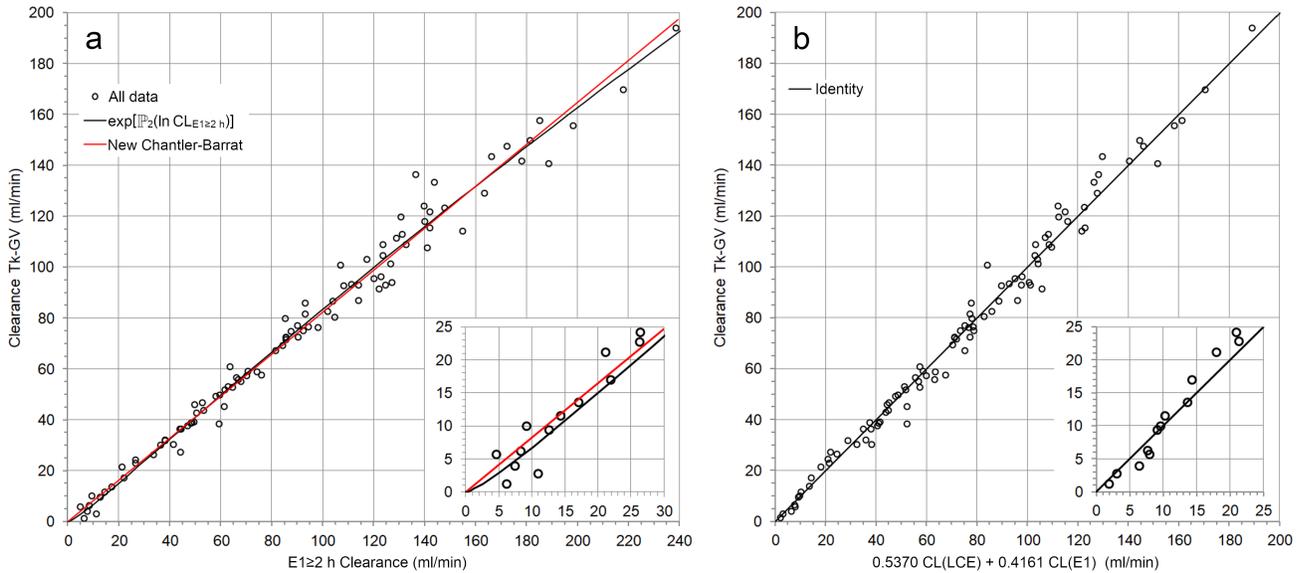} 
\vspace{-1em}
\caption {Shown are 98 cases used with two models to predict Tk-GV GFR results as a reference standard. Panel \textbf{a} shows conversions of E1$\,\geq\,$2 h to Tk-GV clearances using a new Chantler-Barratt slope (red line) and the exponential of a quadratic polynomial of logarithms of E1$\,\geq\,$2 h (black curve). Panel \textbf{b} shows prediction of Tk-GV clearances using the pairwise sum of weighted E1 and LCE CL-values obtained using all samples. }
\label{fig_11}
\vspace{-1em}
\end{figure}

To use Tk-GV as a reference standard for conversion of commonly performed procedures, a new Chantler-Barratt formula was constructed using E1$\,\geq\,$2 h clearance values, see Figure \ref{fig_11}. This yielded,

\begin{equation*}
\begin{aligned}
&\text{CL}_\text{Tk-GV}=0.8243\,\text{CL}_{\text{E1$\geq$\text{2 h}}}\;\\
\text{R}=0.9908,\;\;&\text{SE}=5.89\text{ (ml/min),\;\;\text{Relative error}\,=\,39.0\%} 
\end{aligned}\;\;,
\end{equation*}


\noindent where R is the ANOVA correlation, SE is the standard error of measurements, and the relative error is one standard deviation of proportional error. The new formula yields GFR-values that are $0.8243/0.87=0.9479$ times the old  correction factor's GFR-values because of the new reference standard used, i.e., CL$_{\text{Tk-GV}}$. This provides us with a crude indication of how decreased CL$_{\text{Tk-GV}}$ values are (0.9479$\times$) compared to using corrected CL$_{\text{E1$\geq$\text{2 h}}}$ values. However, the relative error is intractably large, as follows. Less than a raw CL$_{\text{E1$\geq$\text{2 h}}}$ of 60.7 ml/min, which corresponds to a corrected GFR of 50 ml/min, the proportional error is 66.7\%. This is largely due to the 12.5\% of clearances < 50 ml/min that are inflated by 33\% to 307\%. Thus, corrected CL$_{\text{E1$\geq$\text{2 h}}}$-values are not reliable measures of reduced clearance, which provides a justification for clinicians not relying on such measurements. Chantler-Barratt found an actual regression line slope of 0.77 when the line was constrained to go through the origin, and then added 0.10 to make 0.87 as a correction for venous rather than arterial sampling, but did so without any supporting results to validate that hypothesis \cite{Chantler1972}. A finding of 0.82 is the average of 0.77 and 0.87, and is supported by findings. Chantler-Barratt also found problems with nonlinearity especially for low or very high GFR-values. This is not unexpected given that their unconstrained regression line was CL$_{\text{renal}}\approx 0.70\,\text{CL}_{\text{E1}\geq2\text{ h}}+7.2$ (1.73 m$^2\cdot$eBSA$^{-1}\cdot$ml$\cdot$min$^{-1}$). In the \nameref{introduction} section, it was mentioned that a nonlinear correction for E1$\,\geq\,$2 h clearance values should have a slope of zero as it approaches the origin to clear the constant appearing in an unconstrained linear regression. To account for the nonlinearity at the origin, an asymptotically zero slope as CL$\to0$ formula (there are many) was obtained by fitting a quadratic, a ${P}_2(x)=a_0+a_1 x+a_2 \,x^2$, to the logarithms of the 98 Tk-GV and E1$\,\geq\,$2 h clearances provided that $a_2<0$, and as,

\[\mathbb{P}_2(\ln \text{CL}_{\text{Tk-GV}})=-1.088+1.398 \ln \text{CL}_{\text{E1}\geq2\text{ h}}-0.04374\, (\ln \text{CL}_{\text{E1}\geq2\text{ h}})^2\;\;,\]
$a_2=-0.04374<0$, that is indeed the case. To then predict Tk-GV clearances, one takes the exponential of both sides of the above to obtain

\begin{equation*}
\begin{aligned}
\text{CL}_{\text{Tk-GV}}&=\exp\left(-1.088+1.398 \ln \text{CL}_{\text{E1}\geq2\text{ h}}-0.04374 \ln ^2\,\text{CL}_{\text{E1}\geq2\text{ h}}\right)\\
\text{R}&=0.9912,\;\;\text{SE}=5.76\text{ (ml/min),\;\;\text{Relative error}\,=\,27.8\%}
\end{aligned}\;\;,
\end{equation*}
which is improved compared to the errors of the new Chantler-Barratt regression. This is the solid black non-linear curve in Figure \ref{fig_11}a. Notice that in the insert in Figure \ref{fig_11}a that the five smallest GFR values form a pattern that looks like the number 5 on a die, i.e., uncorrelated and disappointingly variable. None of those values are less than 5 ml/min. However, correcting nonlinearity by brute force is unnecessarily complicated. Instead,  Tk-GV CL-values can be estimated by weighted averaging of CL$_\text{LCE}$ and $\text{CL}_{\text{E1}\geq2\text{ h}}$. This represents a weighted average CL-values from the two independent models fit to the same data.

\begin{equation*}
\begin{aligned}
&\text{CL}_{\text{Tk-GV}} = 0.2673\,\text{CL}_{\text{LCE}}+0.6105\, \text{CL}_{\text{E1}\geq2\text{ h}}\\
\text{R}=0.&9919,\;\;\text{SE}=5.53\text{ (ml/min),\;\;\text{Relative error}\,=\,26.6\%}
\end{aligned}\;\;,
\end{equation*}

This equation could be applied to convert E1$\,\geq\,$2 h clearances to approximate Tk-GV clearances. However, the relative error at 26.6\% is still large, and the correction is still suboptimal. No matter how E1$\,\geq\,$2 h CL-values are corrected, the CL-values are too noisy for measuring reduced CL-values to be used for that purpose. This can be improved upon by using methods that do not back extrapolate for 2 h. Note in Figure \ref{fig_7}a that the large magnitude negative slope of an E1 model for increasing renal function is almost perfectly reflected by a strong positive slope for the LCE model. If this is correct, one would expect that some linear combination of E1 and LCE clearances might approximate clearance better than either E1 or LCE taken separately. This is simpler in that theoretically only two samples would be needed for E1-LCE averaging, the potential advantage of which is that similar to a single sample method, one would then need only two sessions with a subject, one just after (flush) bolus intravenous injection of the GFR marker with an early sample drawn at 5 min, and one later. However, the three datasets used here are perhaps not the best ones to explore two plasma sample modelling. Instead, all the samples were included, which yielded,

\begin{equation*}
\begin{aligned}
&\text{CL}_{\text{Tk-GV}}=0.5370\,\text{CL}_{\text{LCE}}+0.4161\,\text{CL}_{\text{E1}}\\
\text{R}=0.9&930,\;\;\text{SE}=5.06 \text{ (ml/min),\;\;\text{Relative error}\,=\,11.7\%}
\end{aligned}\;\;,
\end{equation*}
which although it has slightly better R- and SE-values than correcting E1$\,\geq\,$2 h CL-values, there is a major improvement in relative error. At 11.7\% the relative error, although substantial, is no longer intractably large. To examine how this improvement in relative error has occurred, see the insert in Figure \ref{fig_11}b, which shows that the lower CL-values are now better linearised, and there are two values below 5 ml/min, whereas there were none for estimating Tk-GV CL from E1$\,\geq\,$2 CL-values. Something, then, allowed Tk-GV CL-values to be predicted, without an intercept, whose partial probability, $p=0.3$, indicated discard. If Tk-GV CL-values are reliable, then there should be other ways of predicting them. Indeed, substitution of E2 CL-values for the E1 values above yielded

\begin{equation*}
\begin{aligned}
\text{CL}&_{\text{Tk-GV}}=0.3196\,\text{CL}_{\text{LCE}}+0.6452\,\text{CL}_{\text{E2}}\\
\text{R}=0.9954&,\;\;\text{SE}=4.15\text{ (ml/min),\;\;\text{Relative error}\,=\,10.4\%}
\end{aligned}\;\;,
\end{equation*}
where discard of the insignificant intercept ($-0.60$ ml$\cdot$min$^{-1}$, $p=0.5$) improved the standard error slightly. This then gives us a method of converting E2 model CL-values to Tk-GV CL-values with fairly good precision and accuracy. That such methods exist implies that Tk-GV CL and its V$_\text{MRT}$ investigated above are plausible reference standards. From this latest formula above the 30 estimates that were less than 50 ml/min had a standard error of only 2.13 ml/min but a relative error of 16.4\%. The 67 results > 50 ml/min had a standard error of 4.79 ml/min and a relative error of only 6.17\%. It is important when comparing methods to inspect the range of GFR values being analysed as there are many methods that are unreliable below 50-60 ml/min, for example, CL$_{\text{E1}\geq2\,h}$ as above, and the single sample methods of the literature that use CL$_{\text{E1}\geq2\,\text{h}}$ as a reference standard \cite{Ptacnik2023}.


\subsection*{Limitations}

Long term, steady state, constant infusion renal clearance modelling with bladder catheterisation, e.g. see \cite{Schwartz1949},  can overcome some of the problems associated with bolus, i.e., dynamic, renal clearance. Such data was, unfortunately, not available. Much of the mathematical exploration and statistical testing performed to generate this report have been omitted in order to present the most important observations without undue burden placed upon the reader. For example, the LCE and ln-coth density functions were identified from simplification of a four parameter model and proportional error modelling was selected as best of class from four methods. The regression types tested were ordinary least squares (OLS), 1/y weighted OLS, 1/y$^2$ weighted OLS, and log-log OLS. The Tk-GV model is the only one for which log-log regression is needed (mathematically). All the other regressions presented were 1/y$^2$ weighted OLS. Many formulas, e.g., for constant infusion, half-life of volume and concentration as functions of time were similarly omitted. Alternatives to exponential tails were not extensively tested. Clearance was assumed to be constant in time without proof. The \nameref{sec:appendix} section outlines those derivations specific to this report, where a more complete set of equations is merely a routine application of the calculus to a more complete set of general equations as previously presented and applied respectively for the gamma and gamma-Pareto distributions in \cite{Wesolowski2016PLoS,wesolowski2020comparison}. 

The Tk-GV method has been applied in a clinical setting both retrospectively and prospectively. Four time-samples can be obtained at 5, 20, 60 and 240 min following flush bolus intravenous injection of a good GFR marker. Unlike for E2, nine time-samples obtained up to 8 h post-injection produced CL-values did not significantly differ from the four time-sample, 4-h results when the Tikhonov relative fit error of $\beta$ of a gamma-variate, $K t^{\alpha-1}e^{-\beta\,t}$, was minimised ($n=412$) \cite{wanasundara2016}. For quality assurance, only results for which $\alpha<1$ are correct. In extensive simulation studies using leave-outs, $\alpha>1$ can occur when the first time-sample is obtained later than 10 min or the last sample is obtained earlier than 180 min, this has not occurred clinically. When a saline flush is not used it is not uncommon to create a subcutaneous depot of radioactivity \cite{Wesolowski1988bolus}. In one prospective clinical case a second time-sample was drawn from a partially infiltrated injection site. This led to a spuriously higher concentration at 20 min than at 5 min, and an $\alpha>1$. The incidence of quality assurance failures has been approximately one in 500. 

The LCE and ln-coth models have been presented as fits to multiple samples between approximately 5- to 15-min and 4-, 5- and 24-h. Clinical application should be investigated for more minimalistic curve solutions using only two plasma samples, possibly one at 5- to 15-min and another at 4 h.  However, the three datasets evaluated in this paper are not optimal for such a study and any such modelling is left for future work, as it requires the development a normal range for fractal \cite{west1999fourth} renal metabolic scaling \cite{wesolowski2006improved} with preliminary results suggesting smaller negative fraction and better accuracy for segregating abnormal from normal GFR \cite{Wesolowski2006}.

Scaling of measured GFR is needed to classify sufficiency of renal function versus metabolic demand and should be done with respect to normal measured GFR by 1) normalising powers of variables like volume of distribution and body weight over 2) at least an 8-fold weight range, as well as over 3) a range of abnormal fluid balance, e.g., see \cite{wesolowski2006improved,wesolowski2011validation}. As volume of distribution raised to a power is by far the single most predictive variable for metabolically expected renal function, obtaining volume values uncorrelated to measured and possibly abnormal clearance is highly desirable, but was not determined for the new models of the discussion section. This then would provide for a reference standard for calculating estimating formulas for creatinine, cystatin-C and any other endogenous metabolite. This has not been done in this introductory paper. Clinical correlation, as well as body scaling and normal range calibration are needed for final interpretation of the value of the results.

\section*{Conclusions}\label{sec:conclusions}

The working hypothesis that there are better GFR models than Tk-GV remained unconfirmed. Methods that appear to have potential applicability to the reduced renal function measurement problem are the Tk-GV method and the weighted average of E1 or E2 clearances with LCE clearance values for predicting Tk-GV values. These appear to meet the Schloerb challenge of quantifying anephric conditions. The Tk-GV method produced values that are frequently within the reference standard range, and was the only plasma clearance method tested that was consistently uncorrelated with its weight normalised V$_\text{MRT}$. 

\section*{Acknowledgements}\label{sec:acknowledgements}

The editors and reviewers, especially unnamed Reviewer 1, are thanked for the extensive improvements made during the preparation of this paper. Prof. Geoffrey T. Tucker of the University of Sheffield, Sheffield, UK is thanked for his suggestions concerning this manuscript. Maria T. Burniston and coauthors in the UK \cite{wesolowski2011validation} are thanked for graciously providing Dataset 1. Prof. Jens H. Henriksen of the University of Copenhagen, Denmark is thanked for providing Dataset 2. Prof. Charles D. Russell of the University of Alabama at Birmingham is thanked for providing Dataset 3. Surajith N. Wanasundara is thanked for his help with computer implementation of an earlier version of the Tk-GV processing program.

\section*{Appendix}\label{sec:appendix}

Concentration models have a finite area under the curve (AUC) from $t=0$ to $t=\infty$, i.e., AUC$\,\myeq\int_0^\infty C(t)$. Density functions, $f(t)$, have a total area of one, that is, $\int_0^\infty f(t)=1$, and are found by applying the definition $f(t)=\frac{C(t)}{\text{AUC}}$. Multiplying both sides that definition by AUC and reversing the order of equality yields,

\begin{equation}
\label{eq1}C(t)=\text{AUC }f(t)\;\;,
\end{equation}

\noindent To be clear, AUC is from curve fitting but is the area under the entire curve, not just the data from the first to last time-samples. For example, for E1, let,
$$f(t)=\lambda\, e^{-\lambda\,t},\;\;\;C(t)=\text{AUC}\,\lambda \,e^{-\lambda\,t},$$

\noindent where setting $c=\text{AUC}\,\lambda$ yields the more common notation $C(t)=c \,e^{-\lambda\,t}$. However, $c$ is a dummy variable, i.e., it is unnecessary. In addition to extracting AUC-values immediately from data fitting, identifying the density function makes the rules for its manipulation immediately available. One such rule is the cumulative density function, CDF, also written as $F(t)$, where $F(t) \myeq \int_0^t  f(\tau)\,d\tau$, i.e., the 0 to $t$ integral of $f(t)$, such that $\lim_{t\to\infty}F(t)=1$. The CDF of an exponential density, $\lambda\,e^{-\lambda\,t}$, is thus 

$$F(t)= \int_0^t  \lambda\,e^{-\lambda\,t}\,dx=1-e^{-\lambda\,t}\;\;.$$ 

\noindent As the inside of $-\ln(1-e^{-\beta\,t})$, i.e., $1-e^{-\beta\,t}$, is a cumulative exponential, $-\ln(1-e^{-\beta\,t})$ is a negative Logarithm of a Cumulative Exponential, or LCE as an acronym. To make the LCE into a density function, $-\ln(1-e^{-\beta\,t})$ is multiplied by a constant that makes its total area equal to one, $\lim_{t\to\infty}F(t)=1$. That is,

\begin{equation}\label{eq2}
f(t)=-\frac{6\, \beta }{\pi ^2}\ln \left(1-e^{-\beta\, t}\right)\;\;,
\end{equation}

\noindent where that constant is $\mfrac{6\, \beta }{\pi ^2}$. Combining Eqs.~\eqref{eq1} and \eqref{eq2} yields the fit equation for the LCE model used in this manuscript,

\begin{equation}\label{eq3}
C(t)=\text{AUC}\cdot f(t)=-\text{AUC}\,\frac{6\, \beta }{\pi ^2}\ln \left(1-e^{-\beta\, t}\right)\;\;.
\end{equation}

\noindent \textit{Theorem}. The density function for $-\ln \left(1-e^{-\beta\, t}\right)$ is $f(t)=-\dfrac{6\, \beta }{\pi ^2}\ln \left(1-e^{-\beta\, t}\right)$, i.e., the log cumulative exponential (LCE) distribution. \\\textit{Proof.} We first note that the derivative that yields $-\ln \left(1-e^{-\beta\, t}\right)$ is,

\begin{equation}\label{eqA1}
\frac{d}{d\,t}\left[-\frac{1}{\beta}\text{Li}_2\left(e^{-\beta\, t}\right)\right]=-\ln \left(1-e^{-\beta\, t}\right)\;\;,
\end{equation}
where $\text{Li}_n(z)=\sum _{k=1}^{\infty } \frac{z^k}{k^n}$ is the polylogarithm function of order $n=2.$  $\Big($Hint, let $u=e^{-\beta\,t}$, then $\frac{d}{d\,t}\text{Li}_2\left(e^{-\beta\,t}\right)=\frac{d}{d\,t}\text{Li}_2(u)\frac{d\,u}{d\,t}$, and $\frac{d}{d\,u}\text{Li}_2(u)=-\frac{\ln(1-u)}{u}.\Big)$ Next, we scale $-\ln \left(1-e^{-\beta\, t}\right)$ to be a density function by dividing by the total area from 0 to $t\to \infty$ of its antiderivative. That is since, 
 
\begin{equation}\int_0^{\infty}\frac{d}{d\,t}\left[-\frac{1}{\beta}\text{Li}_2\left(e^{-\beta\, t}\right)\right]=\lim_{t\to \infty}\left[-\frac{1}{\beta}\text{Li}_2\left(e^{-\beta\, t}\right)\right]+\frac{1}{\beta}\text{Li}_2\left(e^{-\beta\cdot 0}\right)=0+\frac{\pi ^2}{6\, \beta}\;\;,\end{equation}

\noindent then,
$$f(t)=-\ln \left(1-e^{-\beta\, t}\right)\bigg/ \frac{\pi ^2}{6\, \beta}=-\frac{6\, \beta }{\pi ^2}\ln \left(1-e^{-\beta\, t}\right)\;\;.\qed$$
\textit{Corollary.} Similarly, the CDF and CCDF = CDF $-$ 1, are from the antiderivative evaluated between 0 and $t$,

\begin{equation}\label{Ft}
F(t)=1-\frac{6 }{\pi ^2}\text{Li}_2\left(e^{-\beta\, t}\right),\;\;\;S(t)=\frac{6 }{\pi ^2}\text{Li}_2\left(e^{-\beta\, t}\right)\;\;,
\end{equation}
where CCDF is the complementary CDF.  The CCDF is symbolised $S(t)$ here, even though  $S(t)$, survival functions, are technically from mass functions, not density functions. For example, the formula for volume of distribution in Table \ref{dists}, V$_{\text{d}}(t)=\text{CL} \frac{1-F(t)}{f(t)}\leftrightarrow \text{Dose} \frac{S(t)}{C(t)}$, i.e., volume of distribution is the surviving dose in the body at a time divided by the concentration at that same time. Note that how long it takes for $F(t)$ to converge to 1 is dependent on a single parameter, $\beta$; the smaller $\beta$ is, the longer it takes. 

The mean residence time, where MRT $=\int_0^\infty t\,f(t)\,dt$ for the LCE density function, was found from evaluating its antiderivative from $t$ equals 0 to $\infty$,

$$\text{MRT}=\frac{6\, \zeta (3)}{\pi ^2\, \beta}\approx\frac{0.730763}{\beta }\;\;,$$
where the zeta ($\zeta$) function of 3 is approximately 1.20206. Note that the ratio of MRT$_\text{LCE}$ and $t_x$ is a constant equal to $\mfrac{6\, \zeta (3)}{\pi ^2\, \Omega}$. That is,  MRT$_\text{LCE}$ occurs at a time approximately 1.2885 times longer than $t_x$, and thus the MRT occurs when the tail is already predominantly an exponential function. 

The \textit{median} residence time ($t_m$, LCE half-survival) was calculated by Newton-Raphson's method for $u$ such that the $S(u)=\frac{6 }{\pi ^2}\text{Li}_2\left(e^{-u}\right)=\frac{1}{2}$. Then, let $u=\beta\,t_{m}$, and solve for $t_{m}$, which yields,

$$t_{m}\approx \frac{0.415389}{\beta}\;\;.$$

\noindent \textit{Theorem}. For ln-coth, $f(t) = \mfrac{4 \beta  }{\pi ^2}\ln \bigg[\coth \left(\mfrac{\beta \, t}{2}\right)\bigg]$ is the density function and the CDF is
$$F(t)=\frac{4 }{\pi ^2}\big[\text{Li}_2(1-y)+\text{Li}_2(-y)+\ln (y+1) \ln (y)\big]+\frac{4}{3},\;\;\text{ where }y=\ln \bigg[\coth \bigg(\frac{\beta \, t}{2}\bigg)\bigg]\;\;.$$
\textit{Proof.} Differentiate. (Hint: $F'(t)=F'(y)\,\dfrac{dy}{dt}$, where $F'(y)=-\dfrac{8 \ln (y)}{\pi ^2 \left(y^2-1\right)}$, $\dfrac{dy}{dt}=-\mfrac{\beta}{2}  \, \text{csch}^2\left(\mfrac{\beta \, t}{2}\right)$, substitute and simplify.) Note that $F(0)=0$ and $\lim_{t\to\infty}F(t)=1$, i.e., $f(t)$ is the ln-coth density function of $F(t).\;\;\qed$

\end{onehalfspacing}

\begin{small}




\begin{thebibliography}{10}
\bibitem{Wilcox1998}
Wilcox R.
\newblock A note on the {T}heil-{S}en regression estimator when the regressor
  is random and the error term is heteroscedastic.
\newblock Biom J. 1998;40(3):261--268.
\end{thebibliography}

\end{small}

\end{document}